\documentclass[journal]{IEEEtran}
\usepackage{amsmath,amssymb,amsfonts,graphicx,mathtools,nccmath,amsthm,float}
\usepackage{algorithm,multicol,multirow,tikz,cite,array,booktabs,url}
\usepackage{algpseudocode,algorithmicx}
\usepackage{cite}
\usepackage{textcomp}
\usepackage{xcolor,flushend}
\usepackage{enumitem}
\usepackage[caption=false]{subfig}
\usepackage[bookmarks,colorlinks]{hyperref}
\usepackage[cal=cm,scr=euler]{mathalpha}

\graphicspath{{figures_Journal/}{./}}
\DeclareGraphicsExtensions{.pdf}

\newcommand{\vect}[1]{\boldsymbol{\mathbf{#1}}}
\newcommand{\norm}[1]{\left\lVert #1 \right\rVert}

\newtheorem{Thm}{Theorem}
\newtheorem{Cor}{Corollary}
\newtheorem{Lem}{Lemma}

\DeclareMathOperator{\trace}{Tr}
\DeclarePairedDelimiter{\abs}{\lvert}{\rvert}

\algnewcommand{\LineComment}[1]{\State \(\#\) #1}
\algnewcommand\algorithmicinput{\textbf{Set}}
\algnewcommand\Set{\item[\algorithmicinput]}
\algnewcommand\algorithmicinitial{\textbf{Initialize}}
\algnewcommand\Initialize{\item[\algorithmicinitial]}

\let\oldReturn\Return
\renewcommand{\Return}{\State\oldReturn}

\begin{document}

\title{Decentralized Cooperative Beamforming for BDRIS-Assisted Cell-Free MIMO OFDM Systems}

\author{Konstantinos D. Katsanos,~\IEEEmembership{Member,~IEEE} and George C. Alexandropoulos,~\IEEEmembership{Senior~Member,~IEEE}
\thanks{This work has been supported by the SNS JU project 6G-DISAC under the EU's Horizon Europe research and innovation program under Grant Agreements number 101139130.}
\thanks{The authors are with the Department of
Informatics and Telecommunications, National and Kapodistrian University of
Athens, 16122 Athens, Greece (e-mails: \{kkatsan, alexandg\}@di.uoa.gr).}
}

\maketitle

\begin{abstract}
In this paper, a wideband cell-free multi-stream multi-user Multiple-Input Multiple-Output (MIMO) Orthogonal Frequency Division Multiplexing (OFDM) system is considered operating within a smart wireless environment enabled by multiple Beyond Diagonal Reconfigurable Intelligent Surfaces (BDRISs). A novel decentralized active and passive beamforming framework, robust to imperfect channel state availability and with minimal cooperation among the system's multiple Base Stations (BSs) for deciding the final configurations of the shared BDRISs, is proposed, which aims to substantially reduce the overhead inherent in centralized solutions necessitating a central processing unit of high computational power. By considering a Dynamic Group-Connected (DGC) BDRIS architecture with frequency-selective responses per unit element, we formulate the system's sum-rate maximization problem with respect to the tunable capacitances and permutation matrices of the BDRISs as well as the precoding matrices of the BSs, which is solved via successive concave approximation and alternating projections as well as consensus-based updates for the BDRISs' design. Through extensive simulation results, it is showcased that the proposed robust decentralized cooperative approach with diverse BDRIS architectures outperforms non-cooperation benchmarks. It is also demonstrated that the considered DGC BDRIS architecture is able to provide sum-rate performance gains sufficiently close to the more complex fully-connected BDRIS structure.
\end{abstract}


\IEEEpeerreviewmaketitle

\section{Introduction} \label{sec:Intro}
The next generations of wireless communication systems are envisioned to support extremely dense device connectivity, substantially higher data throughput, and ultra-low end-to-end latencies, while simultaneously achieving marked enhancements in energy efficiency. Reconfigurable Intelligent Surfaces (RISs) have emerged as a key technology for implementing programmable wireless environments, enabling dynamic control over radio wave propagation~\cite{Basar2024_RIS_6G}. In addition, the cell-free networking paradigm~\cite{ngo2017cell} constitutes a revolutionary architecture envisioning distributed, user-centric cooperation of Base Stations (BSs) that promises enhanced spectral efficiency, coverage uniformity, and interference management, and has been recently explored in combination with RISs~\cite{elhoushy2021cell,zhang2021joint,ma2022cooperative,al2024performance}. 

Cell-free networks typically rely on a Central Processing Unit (CPU) with substantial computational capabilities, which interacts with distributed BSs through dedicated backhaul links~\cite{ngo2017cell}. Each BS is assumed to have perfect local Channel State Information (CSI), which must then be forwarded to the CPU for beamforming design. Consequently, developing active and passive beamforming algorithms for cell-free Multiple-Input Multiple-Output (MIMO) communication systems operating within a smart wireless environment enabled by multiple RISs emerges as a major challenge. Very recently, in \cite{wang2025efficient}, a centralized design accounting for the frequency-selective characteristics of RIS units elements in Orthogonal Frequency Division Multiplexing (OFDM) transmissions was introduced. 

Aiming to reduce the requirement for highly complex CPUs, recent research on cell-free MIMO systems focuses on decentralized schemes~\cite{huang2020decentralized,xu2023algorithm} as well as partially distributed beamforming strategies \cite{ni2022partially}.  In particular, an Alternating Direction Method of Multipliers (ADMM) approach leveraging the random parallel walk technique was developed in~\cite{huang2020decentralized}, while~\cite{xu2023algorithm} proposed a learning-based deep distributed ADMM framework. Both approaches relied on enforcing intricate consensus constraints in scenarios with multiple RISs. On the other hand, to circumvent the complications arising from these constraints, \cite{ni2022partially} introduced a partially distributed architecture in which the CPU is responsible for optimizing the high-dimensional passive beamformer, whereas each BS only performs local optimization of its own low dimensional digital beamformer. However, all above cell-free studies assume perfect global CSI availability collected at a CPU. 

An emerging hardware architecture in the technology of programmable metasurfaces is based on Beyond Diagonal RISs (BDRISs), which realize interconnections between neighboring pairs of response-tunable metamaterials~\cite{maraqa2025beyond,LSN+23}. BDRIS signify a general RIS framework in which the scattering matrix is not constrained to be diagonal~\cite{shen2021modeling}. Consequently, not only the diagonal elements, but also the off-diagonal elements can be flexibly adjusted to control wave propagation~\cite{li2025tutorial}. According to recent studies, BDRISs are able to provide significant performance gains for various important communication-oriented objectives, such as MIMO capacity maximization~\cite{santamaria2024mimo}, sum-rate and energy-efficiency maximization, and power minimization relative to their diagonal counterparts~\cite{zhou2023optimizing}. Moreover, recent investigations have shown that BDRISs are advantageous in cell-free massive MIMO systems as well \cite{hua2025cell,li2025beamforming}. Specifically, in \cite{hua2025cell}, the authors investigate the deployment of a hybrid transmitting–reflecting BDRIS to improve both coverage and spatial multiplexing, while~\cite{li2025beamforming} proposes the integration of a BDRIS to enhance simultaneous wireless information and power transfer. 

In this paper, motivated by the additional design degrees of freedom offered by BDRISs and the expected capacity improvement potential of respective smart wireless environments over uncontrollable propagation scenarios~\cite{JSTSP_distributed_all}, we focus on the distributed optimization of cell-free MIMO systems with Frequency Division Multiplexing (OFDM) transmissions operating within the area of influence of multiple shared frequency-selective BDRISs. The main contributions of this work are summarized as follows.
\begin{itemize}
    \item A novel grouping strategy for the interconnections of the unit elements of frequency-selective BDRISs is presented. In particular, each of the system's BDRISs is realized on the basis of an impedance-switch network, allowing the group indices not restricted to adjacent ones, which provides additional degrees of freedom for optimization for the considered cell-free MIMO OFDM systems.
    \item By adopting the sum-rate maximization metric as the design objective of the considered wideband multi-BDRIS-empowered system, we formulate a challenging non-convex optimization problem with respect to the precoding matrices at each BS as well as the capacitances and permutation matrices of the BDRISs. A novel decentralized design with minimal network-graph-based cooperation enabling consensus on the final configurations of the system's shared BDRIS devices, which is robust to imperfect CSI availability, is presented.  
    \item Extensive simulation results are presented for the proposed distributed design, showcasing the role of the designed minimal cooperation over non-cooperative benchmarks as well as the advantages with the proposed BDRIS architecture over alternative structures. It is also demonstrated that the proposed design is more robust to imperfect CSI cases than a centralized approach, and its sum-rate performance is sufficiently close to CPU-based coordinated designs relying on global CSI availability. 
\end{itemize}
A preliminary version of our decentralized cooperative active-passive beamforming design has recently appeared in~\cite{Katsanos_Cell_Free_conf}, considering, however, conventional diagonal RISs~\cite{Basar2024_RIS_6G} under a cell-free multiple-input single-output system setting. Differently from this work, in this paper, we focus on the more general case of BDRISs and present a novel design for multi-stream MIMO OFDM transmissions.

The remainder of this paper is organized as follows. Section~\ref{sec:Sys_Prob_Form} includes the considered system model as well as the design problem formulation. Section~\ref{sec:Distr_OP_Solution} presents the proposed distributed design framework for sum-rate maximization, while Section~\ref{sec:Numerical} discusses the extensive numerical investigations on the system performance. Finally, Section~\ref{sec:Concl} provides the concluding remarks of the paper. 

\textit{Notations:} Boldface lower-case and upper-case letters represent vectors and matrices, respectively. The transpose, Hermitian transpose, conjugate, inverse, determinant, and the real part of a complex matrix are represented by $\vect{A}^{\rm T}$, $\vect{A}^{\rm H}$, $\vect{A}^*$, $\vect{A}^{-1}$, $|\vect{A}|$, and $\Re\{\vect{A}\}$, respectively. $\mathbb{R}$, $\mathbb{C}$ denote the set of real and complex numbers, respectively, and $\jmath\triangleq\sqrt{-1}$ is the imaginary unit. $\vect{I}_n$ ($n\geq2$) is the $n\times n$ identity matrix, $\vect{e}_{n'}$ is the $n'$-th column of $\vect{I}_n$ with $n'=1,2,\dots,n$, $\vect{0}_{m\times n}$ is the $m\times n$ zeros' matrix, and $\vect{1}_m$ is the all-ones vector with $m$ entries. $\trace(\vect{A})$ and $\norm{\vect{A}}_{\rm F}$ represent $\vect{A}$'s trace and its Frobenius norm, respectively, while $<\vect{A}_1,\vect{A}_2>\triangleq\Re\{\trace(\vect{A}_1^H\vect{A}_2)\}$ is the inner product between the matrices $\vect{A}_1$ and $\vect{A}_2$ of suitable dimensions and $\vect{A}_1\otimes\vect{A}_2$ represents the Kronecker product. $\mathbb{E}[\cdot]$ denotes the expectation operator and $\mathbf{x}\sim\mathcal{CN}(\mathbf{a},\mathbf{A})$ indicates a complex Gaussian random vector with mean $\vect{a}$ and covariance matrix $\mathbf{A}$. $[\vect{A}]_{i,j}$ represents the $(i,j)$-th element of matrix $\vect{A}$, $[\vect{A}]_{i:j,:}$ ($[\vect{A}]_{:,i:j}$) is a sub-matrix formed by $\vect{A}$'s entries from $i$ to $j$ rows (columns), and $\operatorname{vec}(\vect{A})$ stands for the operator obtained by stacking column-wise $\vect{A}$'s elements, with $\operatorname{unvec}(\vect{a})$ indicating the inverse operation. $\vect{A}_1\bigoplus\vect{A}_2$ denotes the direct sum between two square matrices $\vect{A}_1$ and $\vect{A}_2$, which is a block diagonal matrix whose non-zero diagonal blocks are $\vect{A}_1$ and $\vect{A}_2$; $\bigoplus_{i=1}^n\vect{A}_i$ is the extension of this operation to an arbitrary, but finite, number of matrices. $\nabla_{\mathbf{A}}f$ denotes the gradient vector of the function $f(\cdot)$ along the direction indicated by $\mathbf{A}$. Finally, $\mathcal{O}(\cdot)$ represents the Big-O notation. %for the function $f(x)$. 

%%%%%%%%%%%%%%%%%%%%%%%%%%%%%%%%%%%%%%%%%%%%%%%%%%%%%%%%
%                   Section Change                     %
%%%%%%%%%%%%%%%%%%%%%%%%%%%%%%%%%%%%%%%%%%%%%%%%%%%%%%%%
\section{System Model and Problem Formulation} \label{sec:Sys_Prob_Form}
This section outlines the considered system model comprising multiple BSs, reconfigurable metasurfaces, and Users' Equipment (UEs) which are deployed in a cell-free manner, together with the proposed BDRIS architecture and the resulting received signal model. The joint design objective for the BSs precoders and the tunable parameters of the multiple BDRISs, shared among BSs and UEs, is also introduced. 

\subsection{System Model} \label{sec:Sys_Model}
A cell-free MIMO wireless system consisting of $B$~BSs and $R$ BDRISs and performing data communication in the downlink direction towards $U$ UEs is considered. Each $b$-th BS ($b=1,2,\ldots,B$) is equipped with $N_t$ antenna elements and performs OFDM transmissions over $K$ Sub-Carriers (SCs), whereas each $u$-th UE ($u=1,2,\ldots,U$) and $r$-th BDRIS ($r=1,2,\ldots,R$) possess $N_r$ antennas and $M$ unit elements of reconfigurable electromagnetic responses, respectively. Let $\vect{x}_{b,k} \triangleq \sum_{u=1}^U \vect{W}_{b,u,k}\vect{s}_{u,k}$ represent the transmitted signal from each $b$-th BS, where $\vect{W}_{b,u,k}\in\mathbb{C}^{N_t\times N_s}$ denotes the digital precoding matrix intended for the data symbol $\vect{s}_{u,k} \sim \mathcal{CN}(\vect{0}_{N_s},\vect{I}_{N_s})$ with $N_s \leq \min\{N_r,N_t\}$ (usually chosen from a discrete modulation set), intended for the $u$-th UE at the $k$-th SC. Under the assumption that each $b$-th BS is capable of transmitting with maximum power equal to $P_b^{\max}$, the constraint $\sum_{u=1}^U\sum_{k=1}^K \norm{\vect{W}_{b,u,k}}_{\rm F}^2\leq P_b^{\max}$ must be satisfied.

In the considered cell-free MIMO system, the deployed BDRISs can be shared among all BSs and UEs, and thus need to be jointly designed by the formers~\cite{Alexandropoulos2023_RISDeployment}. This is in large contrast with the state-of-the-art multi-RIS-empowered system designs where each metasurface is exclusively controlled and optimized by a distinct BS~\cite{JSTSP_distributed,gao2025multi,lim2025distributed}, even if each of them might impact nearby BSs (e.g., when installed at the cell boundaries in cellular system setups). In addition, typical cell-free network architectures necessitate the adoption of a CPU of high computational power, being responsible to interact to each BS, and, if present, to each RIS controller, through dedicated control links. This unit actually collects all required inputs (e.g., instantaneous CSI) to formulate the overall system design objective, which is then solved. The  optimized system parameters are finally transferred to the respective system entities (e.g., BS precoding matrices and RIS configurations). As will be detailed in the sequel, the proposed cell-free MIMO system operates in a decentralized manner with minimal cooperation to enable consensus on the configurations of the system's shared BDRIS devices. It is assumed that each BS estimates all instantaneous CSI involving it, which is then shared among the rest of the BSs via dedicated error-free control links. The same happens for the instantaneous CSI concerning BDRIS channels, which can be first collected at predefined BSs (examples of CSI estimation techniques for RIS-based systems are available in~\cite{Jian2022_RIS_Hardware,HRIS_CE,Swindlehurst_CE}) and then shared among them. Then, each BS solves an optimization problem to design its multi-UE precoding matrix as well as the configurations of the BDRISs. For the latter design, relevant parameter sharing takes place among the BSs during the optimization process via the dedicated error-free control links, with the goal to lead to consensus on the final configurations of all shared BDRISs in the cell-free MIMO OFDM system.  

\subsection{BDRIS Architecture and Frequency-Dependent Response} \label{sec:RIS_Response}
A Dynamic Group-Connected (DGC) architecture is considered for each $r$-th BDRIS~\cite{li2023dynamic}, according to which the number of their elements $M$ is divided into $G$ independent groups, where the number of elements per group is equal to $\tilde{M} \triangleq M/G$. In particular, the BDRIS configuration matrix $\vect{\Phi}_{r,k}$ has the following block diagonal structure: $\vect{\Phi}_{r,k} = \bigoplus_{g=1}^G \vect{\Phi}_{r,k,g}$ with $\vect{\Phi}_{r,k,g}\in\mathbb{C}^{\tilde{M}\times \tilde{M}}$. On the other hand, the grouping strategy is a design parameter represented by the matrix $\vect{Q}_{p,r}\in\mathbb{R}^{M\times M}$, which can be practically implemented with an impedance-switch network~\cite{li2023dynamic}. In particular, $\vect{Q}_{p,r}$ is a permutation matrix whose role is to rearrange the sequential order of the common Group-Connected (GC) architecture, so that each $g$-th group $\mathcal{G}_g$ ($g=1,2,\dots,G$) is composed of the elements indexed by \cite{nerini2024static} yielding: 
\begin{equation}
    \mathcal{G}_g \triangleq \left\{ p\left((g-1)\tilde{M} + 1\right), \dots, p\left(g \tilde{M}\right) \right\},
\end{equation}
and thus $\vect{Q}_{p,r} = [\vect{e}_{p(1)},\dots,\mathbf{e}_{p(M)}]$. Without loss of generality, it is also assumed that the proposed BDRIS architecture is reciprocal, which means that the overall response matrix of the metasurfaces, $\widetilde{\vect{\Phi}}_{r,k} \in \mathbb{C}^{M\times M}$, obeys the expression:
\begin{equation} \label{eqn:Phi_complete}
    \widetilde{\vect{\Phi}}_{r,k} = \vect{Q}_{p,r} \vect{\Phi}_{r,k} \vect{Q}_{p,r}^{\rm T}\, \forall r\in\mathscr{R}, \forall k\in\mathscr{K},
\end{equation}
with $\mathscr{R}\triangleq\{1,2,\ldots,R\}$ and $\mathscr{K}\triangleq\{1,2,\dots,K\}$, where $\vect{\Phi}_{r,k}$ needs to satisfy the reciprocity condition $\vect{\Phi}_{r,k}^{\rm T} = \vect{\Phi}_{r,k}$, as well as the constraint $\vect{\Phi}_{r,k} \vect{\Phi}_{r,k}^{\rm H} \preceq \vect{I}_M$ for energy conservation. 

We further adopt the general assumption that each $n$-th unit element ($n=1,2,\dots,\tilde{M}$) of each $g$-th group in each $r$-th BDRIS can be characterized as an equivalent resonant circuit comprising a resistor $R_0$, a tunable capacitor $[\vect{C}_{r,g}]_{n,n'}$ ($n'=1,2,\dots,\tilde{M}$), an inductor $L_2$ connected in series, and one more inductor $L_1$ connected in parallel with the first three circuit elements for the interconnections between the elements $n$ and $n'$ (with $n\neq n'$) of the $g$-th group. Then, each fully-connected group of each $r$-th BDRIS can be treated as a multiport network whose response configuration can be modeled by the scattering matrix $\vect{\Phi}_{r,k,g}$ that can be expressed in the frequency domain as follows~\cite{nerini2024universal}:
\begin{equation} \label{eqn:RIS_Scattering}
    \vect{\Phi}_{r,k,g}(\vect{C}_{r,g},f_k) = (\vect{A}_{r,k,g} + \psi_0\vect{I}_{\tilde{M}})^{-1}(\vect{A}_{r,k,g} - \psi_0\vect{I}_{\tilde{M}}),
\end{equation}
where $f_k$ denotes the $k$-th SC frequency and $\psi_0$ is the characteristic admittance of the propagation medium, while $\vect{A}_{r,k,g} \equiv \vect{A}_{r,k,g}(\vect{C}_{r,g},f_k) \in \mathbb{C}^{\tilde{M}\times\tilde{M}}$ is a matrix that collects the characteristic admittances of the equivalent circuits between the connections of all the $r$-th BDRIS elements. In particular, it is assumed that $\vect{C}_{r,g} \in \mathbb{R}^{\tilde{M}\times\tilde{M}}$ is a symmetric matrix, while the entries of $\vect{A}_{r,k,g}$ are given for $\kappa\triangleq 2\pi$ by:
\begin{equation} \label{eqn:impedances}
\begin{aligned}
&[\vect{A}_{r,k,g}(\vect{C}_{r,g},f_k)]_{n,n'} = \\
&\begin{cases}
\displaystyle
\sum\limits_{\bar{n}=1}^{\tilde{M}} \left( \frac{1}{R_0 + \jmath\kappa f_k L_2 + \frac{1}{\jmath\kappa f_k [\vect{C}_{r,g}]_{n,\bar{n}}}} + \frac{1}{\jmath\kappa f_k L_1} \right), & n = n', \\
\displaystyle
-\left( \frac{1}{R_0 + \jmath\kappa f_k L_2 + \frac{1}{\jmath\kappa f_k [\vect{C}_{r,g}]_{n,n'}}} + \frac{1}{\jmath\kappa f_k L_1} \right), & n \neq n'.
\end{cases}
\end{aligned}
\end{equation}
It is noted that the symmetry of the matrix $\vect{C}_{r,g}$ guarantees the symmetry of $\vect{\Phi}_{r,k,g}$ and thus of $\vect{\Phi}_{r,k}$.

\subsection{Received Signal Model} \label{sec:Signal_Model}
According to the considered cell-free network architecture, there are in total $(R+1)BU$ communicating BS-UE links, among which $BU$ are direct links between BSs and UEs, while the rest are associated with the BDRIS-assisted tunable reflections. Let $\vect{H}_{b,u,k} \in \mathbb{C}^{N_t\times N_r}$ denote the direct channel gain matrix between the $b$-th BS and the $u$-th UE at each $k$-th SC. There are also BS-BDRIS-UE links through which the signals transmitted by the $b$-th BS are reflected by the $r$-th BDRIS before arriving at the $u$-th UE. Let $\vect{F}_{b,r,k} \in \mathbb{C}^{M\times N_t}$ and $\vect{G}_{r,u,k}\in\mathbb{C}^{N_r\times M}$ represent each BS-BDRIS channel and each BDRIS-UE channel, respectively, for each $k$-th SC. Then, the baseband received signal at each $u$-th UE can be mathematically expressed in the frequency domain as follows:
\begin{equation} \label{eqn:overall_received_signal}
    \vect{y}_{u,k} \triangleq \sum_{b=1}^B \vect{y}_{b,u,k} + \vect{n}_{u,k},
\end{equation}
where $\vect{y}_{b,u,k}$ denotes the signal received from each $b$-th BS, which is given by the expression:
\begin{equation} \label{eqn:partial_received_signal}
   \vect{y}_{b,u,k} \triangleq \left(\vect{H}_{b,u,k} + \sum_{r=1}^R \vect{G}_{r,u,k}\widetilde{\vect{\Phi}}_{r,k}\vect{F}_{b,r,k}\right)\sum_{q=1}^U\vect{W}_{b,q,k}\vect{s}_{q,k},
\end{equation}
where $\vect{n}_{u,k}\sim\mathcal{CN}(\vect{0},\sigma_{u,k}^2\vect{I}_{N_r})$ represents the Additive White Gaussian Noise (AWGN) modeling the thermal noise at each $u$-th UE receiver. Using the definition $\widetilde{\vect{H}}_{b,u,k}\triangleq \vect{H}_{b,u,k} + \sum_{r=1}^R \vect{G}_{r,u,k}\widetilde{\vect{\Phi}}_{r,k}\vect{F}_{b,r,k} \in \mathbb{C}^{N_t\times N_r}$, the received signal model in \eqref{eqn:overall_received_signal}, can be rewritten more compactly as follows:
\begin{equation} \label{eqn:compact_received_signal}
\begin{aligned}\vect{y}_{u,k}=&\sum_{b=1}^B\widetilde{\vect{H}}_{b,u,k}\vect{W}_{b,u,k}\vect{s}_{u,k}\\
&+\sum_{\substack{q=1,\\q\neq u}}^U\sum_{b=1}^B \widetilde{\vect{H}}_{b,u,k}\vect{W}_{b,q,k}\vect{s}_{q,k} + \vect{n}_{u,k},
\end{aligned}
\end{equation}
where the first summation is the desired signal at each $u$-th UE, while the second set of summation terms stands for the total interference signal. 

\subsection{System Design Objective} \label{sec:Prob_Form}
Based on the received signal model in \eqref{eqn:compact_received_signal}, the following matrices with system's design parameters are defined: \textit{i}) the precoding matrix $\widetilde{\vect{W}}\triangleq \left[\tilde{\vect{W}}_1^{\rm T},\dots,\tilde{\vect{W}}_B^{\rm T}\right]^{\rm T} \in \mathbb{C}^{BUKN_t\times N_s}$, with $\tilde{\vect{W}}_b\triangleq \left[\vect{W}_{b,1}^{\rm T},\dots,\vect{W}_{b,U}^{\rm T}\right]^{\rm T}\in \mathbb{C}^{UKN_t\times N_s}$ and $\vect{W}_{b,u}\triangleq \left[\vect{W}_{b,u,1}^{\rm T},\dots,\vect{W}_{b,u,K}^{\rm T}\right]^{\rm T} \in \mathbb{C}^{KN_t\times N_s}$; \textit{ii}) the tunable capacitors $\widetilde{\vect{C}}\triangleq \bigoplus_{r=1}^R \vect{C}_r\in \mathbb{R}^{RM\times RM}$; and \textit{iii}) the permutation matrix $\widetilde{\vect{Q}}_p \triangleq \bigoplus_{r=1}^R \vect{Q}_{p,r} \in \mathbb{R}^{RM\times RM}$. By treating multi-UE interference as an additional source of noise at each $u$-th UE side, the achievable instantaneous rate performance in bits per second per Hz (bits/s/Hz) for each $u$-th UE can be expressed as the following function of the system design variables (after omitting the constant multiplicative factor $1/K$):
\begin{equation} \label{eqn:sum_rate_per_user}    \mathcal{R}_u\left(\widetilde{\vect{W}},\widetilde{\vect{C}},\widetilde{\vect{Q}}_p\right) = \sum_{k=1}^K\log_2\left|\vect{I}_{N_s} + \vect{S}_{u,u,k}^{\rm H} \vect{P}_{u,k}^{-1} \vect{S}_{u,u,k}\right|,
\end{equation}
where $\vect{P}_{u,k} \triangleq \sum_{q=1,q\neq u}^U \vect{S}_{u,q,k} \vect{S}_{u,q,k}^{\rm H} + \sigma_{u,k}^2\vect{I}_{N_r}$ and $\vect{S}_{u,q,k} \triangleq \sum_{b=1}^B \widetilde{\vect{H}}_{b,u,k} \vect{W}_{b,q,k} \in \mathbb{C}^{N_r \times N_s}$. In this paper, we focus on maximizing the achievable sum rate of the considered wideband cell-free multi-BDRIS-empowered MIMO system under the assumption that imperfect instantaneous CSI knowledge is available. In mathematical terms, this paper's system design problem is formulated as follows:
\begin{align*}
	\mathcal{OP}: \,\max_{\widetilde{\vect{W}},\widetilde{\vect{C}},\widetilde{\vect{Q}}_p} \, & \quad \mathcal{R} \triangleq \mathbb{E}_{\vect{H},\vect{G},\vect{F}} \left[\sum_{u=1}^U \mathcal{R}_u\left(\widetilde{\vect{W}},\widetilde{\vect{C}},\widetilde{\vect{Q}}_p\right)\right] \\
	\text{s.t.} & \quad \sum_{u=1}^{U}\sum_{k=1}^{K} \left\|\vect{W}_{b,u,k}\right\|_{\rm F}^2 \leq P_b^{\max} \;\;\forall b \in \mathscr{B},\\
    & \quad \vect{C}_{r,g} = \vect{C}_{r,g}^{\rm T}\;\; \forall r \in \mathscr{R}, g \in \mathscr{G}, \\
	& \quad C_{\min} \leq [\vect{C}_{r,g}]_{n,n'} \leq C_{\max}\;\; \forall n,n'\in\mathscr{N},\\
    & \quad \vect{Q}_{p,r} \in \mathcal{Q}\;\; \forall r \in \mathscr{R},
\end{align*}
where the expectation in the objective function is taken with respect to the estimation errors for all channel gains encapsulated into $\widetilde{\vect{H}}_{b,u,k}$ for all indices $b$, $u$, $k$, as well as set $\mathscr{B}\triangleq\{1,2,\ldots,B\}$, $\mathscr{G} \triangleq\{1,2,\dots,G\}$, and $\mathscr{N}\triangleq\{1,2,\dots,\tilde{M}\}$. Finally, parameters $C_{\min}$ and $C_{\max}$ in the second constraint denote the permissible minimum and maximum values for the BDRIS tunable capacitors as determined by circuit characteristics, while $\mathcal{Q}\triangleq \{\vect{Q}_{p,r} \in \{0,1\}^{M \times M}: \vect{Q}_{p,r}\vect{1}_M = \vect{1}_M,\vect{Q}_{p,r}^{\rm T}\vect{1}_M=\vect{1}_M \}$ indicates the feasible set for the permutation matrices at each DGC BDRIS. 

\section{Distributed Sum-Rate Maximization} \label{sec:Distr_OP_Solution}
In this section, we focus on solving $\mathcal{OP}$ based on the decentralized cooperative design framework introduced in Section~\ref{sec:Sys_Model}. In particular, the multiple BSs are treated as agents whose objective is to maximize the sum-rate function $\mathcal{R}$ cooperatively, with minimum exchange messages among them. However, in contrast to the cellular case \cite{JSTSP_distributed,IBC_distributed2024}, $\mathcal{OP}$'s objective function cannot be decomposed on a per agent basis, a fact that induces inherent difficulties in distributively handling it. Nevertheless, we next propose a decentralized optimization framework that takes under consideration that all BDRISs can be shared among all BSs and UEs. In particular, the goal is, for all BSs, to agree on a common set of the BDRIS configuration parameters $\widetilde{\vect{C}}$ and $\widetilde{\vect{Q}}_p$, while the BS precoding matrices included in $\tilde{\vect{W}}_b$ are optimized individually. Consequently, each BS possesses its own set of variables, among which the BDRISs' involved need to be common, or, in other words, the BSs need to achieve consensus on them during the design optimization process.

Based on the above observations and capitalizing on the stochastic decomposition method detailed in \cite{yang2016parallel}, the distributed version of $\mathcal{OP}$, aimed to be solved by each $b$-th BS, can be mathematically expressed as follows:
\begin{align*}
	\mathcal{OP}_{\mathcal{D},b}: \,\max_{\vect{X}_b \in \mathcal{X}} \, \quad \sum_{u=1}^U\widehat{\mathcal{R}}_u\left(\vect{X}_b;\vect{X}_{-b}^t,\vect{\eta}_b^t\right),
\end{align*}
where $\vect{X}_b \triangleq \left[\tilde{\vect{W}}_b^{\rm T},\widetilde{\vect{C}}_b^{\rm T},\widetilde{\vect{Q}}_{p,b}^{\rm T}\right]^{\rm T}$ includes the design parameters for the $b$-th BS with $\mathcal{X}$ denoting the feasible set defined as the intersection of the constraints set stated in $\mathcal{OP}$, while, for each $t$-th algorithmic iteration, $\vect{X}_{-b}^t$ collects all other BSs' parameters and $\vect{\eta}_b^t$ represents the locally generated noisy channel samples. In addition, the $u$-th term in the objective function of $\mathcal{OP}_{\mathcal{D},b}$ is given by expression:
\begin{equation}
\begin{aligned}
    \widehat{\mathcal{R}}_u(\vect{X}_b;\vect{X}_{-b}^t,\vect{\eta}_b^t) \triangleq &\,\rho^t\widetilde{\mathcal{R}}_u(\vect{X}_b;\vect{X}_{-b}^t,\vect{\eta}_b^t) \\
    &+ \rho^t<\vect{\Pi}_{\vect{X}_b}^t,\vect{X}_b - \vect{X}_b^t> \\
    &+ (1- \rho^t)<\vect{D}_{\vect{X}_b}^t,\vect{X}_b - \vect{X}_b^t> \\
    &- \frac{\tau}{2}\norm{\vect{X}_b - \vect{X}_b^t}_{\rm F}^2,
\end{aligned}
\end{equation}
where the first term represents the surrogate function of $\mathcal{R}_u$, which is defined as follows: 
\begin{equation}
\begin{aligned}
&\widetilde{\mathcal{R}}_u(\vect{X}_b;\vect{X}_{-b}^t,\vect{\eta}_b^t) \triangleq \mathcal{R}_u(\vect{X}_b;\vect{X}_{-b}^t,\vect{\eta}_b^t)\\
&\qquad+ <\vect{\Gamma}_{\widetilde{\vect{C}}_b}^t,\widetilde{\vect{C}}_b - \widetilde{\vect{C}}_b^t> + <\vect{\Gamma}_{\widetilde{\vect{Q}}_{p,b}}^t,\widetilde{\vect{Q}}_{p,b} - \widetilde{\vect{Q}}_{p,b}^t>,
\end{aligned}
\end{equation}
with $\vect{\Gamma}_{\widetilde{\vect{C}}_b}^t \triangleq \nabla_{\widetilde{\vect{C}}_b}\mathcal{R}_u(\widetilde{\vect{C}}_b;\vect{X}_{-b}^t,\vect{\eta}_b^t)\Big\vert_{\widetilde{\vect{C}}_b=\widetilde{\vect{C}}_b^t}$, $\vect{\Gamma}_{\widetilde{\vect{Q}}_{p,b}}^t \triangleq \nabla_{\widetilde{\vect{Q}}_{p,b}}\mathcal{R}_u(\widetilde{\vect{Q}}_b;\vect{X}_{-b}^t,\vect{\eta}_b^t)\Big\vert_{\widetilde{\vect{Q}}_b = \widetilde{\vect{Q}}_b^t}$ having a role in the decoupling of $\tilde{\vect{W}}_b$, $\widetilde{\vect{C}}_b$, and $\widetilde{\vect{Q}}_{p,b}$ in the objective function $\mathcal{R}_u$. In addition, $\vect{\Pi}_{\vect{X}_b}^t \triangleq \sum_{q=1,q\neq u}^U \nabla_{\vect{X}_b}\mathcal{R}_q(\vect{X}_b;\vect{X}_{-b}^t,\vect{\eta}_b^t)\Big\vert_{\vect{X}_b=\vect{X}_b^t}$ is characterized as the pricing matrix \cite{JSTSP_distributed}, while $\vect{D}_{\vect{X}_b}^t$ serves as an online estimate of the gradient of the objective function at the realization $\vect{\eta}_b^t$, which is updated recursively as\footnote{We note that, from this point on, we are about to slightly abuse the notation $\nabla_{\vect{x}}f(\vect{x})\vert_{\vect{x}=\vect{x}^t}$ and replace it instead with $\nabla_{\vect{x}}f(\vect{x}^t)$ for brevity.} \cite{yang2016parallel}:
\begin{equation} \label{eqn:accum_definition}
    \vect{D}_{\vect{X}_b}^t = (1- \rho^t)\vect{D}_{\vect{X}_b}^{t-1} + \rho^t\left( \vect{\Pi}_{\vect{X}_b}^t + \nabla_{\vect{X}_b}\mathcal{R}_u(\vect{X}_b^t;\vect{X}_{-b}^t,\vect{\eta}_b^t) \right),
\end{equation}
with $\rho^t\in (0,1]$ being a suitable step-size sequence (letting $\rho^0=1$) and $\tau > 0$ to ensure strong concavity of the surrogate function $\widehat{\mathcal{R}}_u(\vect{X}_b;\vect{X}_{-b}^t,\vect{\eta}_b^t)$. In the next subsections, we solve $\mathcal{OP}_{\mathcal{D},b}$ locally for each of the three sets of design parameters included in $\vect{X}_b$, that is, $\tilde{\vect{W}}_b$, $\widetilde{\vect{C}}_b$ and $\widetilde{\vect{Q}}_{p,b}$. 

\subsection{Precoding Optimization at Each $b$-th BS} \label{sec:Precoder_Design}
The optimization problem with respect to the precoding matrix of the $b$-th BS reduces to the following formulation:
\begin{align*}
    \mathcal{OP}_{\tilde{\vect{W}}_b}: \,\max_{\vect{W}_{b,u,k}} \, & \quad f(\vect{W}_{b,u,k})\triangleq\rho^t\mathcal{R}_{u,k}(\vect{W}_{b,u,k};\vect{X}_{-b}^t,\vect{\eta}_b^t) \\
    &+ \rho^t<\vect{\Pi}_{\vect{W}_{b,u,k}}^t,\vect{W}_{b,u,k}-\vect{W}_{b,u,k}^t> \\
    &+(1-\rho^t)<\vect{D}_{\vect{W}_{b,u,k}}^{t},\vect{W}_{b,u,k} - \vect{W}_{b,u,k}^t> \\
    &- \frac{\tau}{2}\norm{\vect{W}_{b,u,k} - \vect{W}_{b,u,k}^t}_{\rm F}^2 \\
	\text{s.t.} & \quad \sum_{u=1}^{U}\sum_{k=1}^{K} \norm{\vect{W}_{b,u,k}}_{\rm F}^2 \leq P_b^{\max}.
\end{align*}
To proceed with the solution of this problem, we first deal with the non-concavity of the first term in its objective function (i.e., the logarithmic term $\mathcal{R}_{u,k}$), for which we make use of the following lemma. 
\begin{Lem} \label{lem:surrogate_precoder}
    Let the following matrix definitions: $\vect{T}_{u,k} \triangleq \vect{S}_{u,u,k}\vect{S}_{u,u,k}^{\rm H} + \vect{P}_{u,k}$, $\vect{L}_{u,k} \triangleq \vect{I}_{N_s} - \vect{S}_{u,u,k}^{\rm H}\vect{T}_{u,k}^{-1}\vect{S}_{u,u,k}$, $\vect{M}_{u,k} \triangleq \vect{L}_{u,k}^{-1} \vect{S}_{u,u,k}^{\rm H} \vect{T}_{u,k}^{-1}$, $\vect{N}_{u,k} \triangleq \vect{T}_{u,k}^{-1}\vect{S}_{u,u,k}\vect{M}_{u,k}$, and $\vect{R}_{u,k} \triangleq \sum_{b'=1,b'\neq b}^B \widetilde{\vect{H}}_{b',u,k}\vect{W}_{b',u,k}$. Then, $\mathcal{R}_{u,k}$, i.e., the $k$-th summand of $\mathcal{R}_u$ expressed in \eqref{eqn:sum_rate_per_user}, can be lower-bounded by the surrogate function $\breve{\mathcal{R}}_{u,k}(\vect{W}_{b,u,k})$. In fact, it holds that:
    \begin{align*}
        \mathcal{R}_{u,k} &\geq \breve{\mathcal{R}}_{u,k}(\vect{W}_{b,u,k}) \\
        &= \frac{-1}{\ln(2)}\Bigg( \trace\left(\widetilde{\vect{H}}_{b,u,k}^{\rm H}\vect{N}_{u,k}\widetilde{\vect{H}}_{b,u,k}\vect{W}_{b,u,k}\vect{W}_{b,u,k}^{\rm H}\right) \\
        &\qquad\qquad+ \trace\left(\vect{R}_{u,k}^{\rm H}\vect{N}_{u,k}\widetilde{\vect{H}}_{b,u,k}\vect{W}_{b,u,k}\right) \\
        &\qquad\qquad+ \trace\left(\widetilde{\vect{H}}_{b,u,k}^{\rm H}\vect{N}_{u,k}\vect{R}_{u,k}\vect{W}_{b,u,k}^{\rm H}\right) \\
        &\qquad\qquad- 2\Re\left\{\trace\left(\vect{M}_{u,k}\widetilde{\vect{H}}_{b,u,k}\vect{W}_{b,u,k}\right)\right\} \Bigg).
    \end{align*}
\end{Lem}
\begin{IEEEproof}
    The proof is delegated in Appendix~\ref{apx:surrogate_precoder_proof}.
\end{IEEEproof}

Before we present the solution to $\mathcal{OP}_{\tilde{\vect{W}}_b}$, we first elaborate on the expressions for $\vect{\Pi}_{\vect{W}_{b,u,k}}^t$ and $\vect{D}_{\vect{W}_{b,u,k}}^{t}$, that can be derived by computing the gradient expressions according to their definitions. In particular, if we let $\vect{K}_{q,k} \triangleq \vect{I}_{N_s} + \vect{S}_{q,q,k}^{\rm H}\vect{P}_{q,k}^{-1}\vect{S}_{q,q,k}$, it can be shown that the desired expression for the pricing matrix is given by:
\begin{equation} \label{eqn:pricing_precoder}
    \vect{\Pi}_{\vect{W}_{b,u,k}}^t = \frac{-1}{\ln(2)}\sum_{\substack{q=1,\\q\neq u}}^U \widetilde{\vect{H}}_{b,q,k}^{\rm H} \vect{P}_{q,k}^{-1} \vect{S}_{q,q,k} \vect{K}_{q,k}^{-1} \vect{S}_{q,q,k}^{\rm H} \vect{P}_{q,k}^{-1} \vect{S}_{q,u,k},   
\end{equation}
while for $\vect{D}_{\vect{W}_{b,u,k}}$, it suffices to compute $\nabla_{\vect{W}_{b,u,k}}\mathcal{R}_{u,k}$ according to \eqref{eqn:accum_definition}, whose expression is derived as:
\begin{equation} \label{eqn:gradient_W_Accum}
    \nabla_{\vect{W}_{b,u,k}}\mathcal{R}_{u,k} = \frac{1}{\ln(2)}  \widetilde{\vect{H}}_{b,u,k}^{\rm H}\vect{P}_{u,k}^{-1}\vect{S}_{u,u,k}\vect{K}_{u,k}^{-1}.
\end{equation}
Hence, the combination of \eqref{eqn:pricing_precoder} and \eqref{eqn:gradient_W_Accum} yields $\vect{D}_{\vect{W}_{b,u,k}}^t$. 

To proceed with the solution of $\mathcal{OP}_{\tilde{\vect{W}}_b}$, it suffices to replace $\mathcal{R}_{u,k}$ by $\breve{\mathcal{R}}_{u,k}$ according to Lemma~\ref{lem:surrogate_precoder}, and note that it is reduced to a concave problem. Hence, it can be readily solved with its Lagrangian function, whose expression is given by:
\begin{align*}
    &\mathcal{L}(\vect{W}_{b,u,k},\lambda) = \rho^t\breve{\mathcal{R}}_{u,k}(\vect{W}_{b,u,k}) \\ 
    &\qquad+ \rho^t\Re\left\{\trace\left(\vect{\Pi}_{\vect{W}_{b,u,k}}^{t,{\rm H}}\left(\vect{W}_{b,u,k} - \vect{W}_{b,u,k}^t\right)\right) \right\} \\
    &\qquad+ (1-\rho^t)\Re\left\{\trace\left( \vect{D}_{\vect{W}_{b,u,k}}^{t,{\rm H}}\left(\vect{W}_{b,u,k} - \vect{W}_{b,u,k}^t\right) \right) \right\} \\
    &\qquad- \frac{\tau}{2}\trace\left( \left( \vect{W}_{b,u,k} -\vect{W}_{b,u,k}^t \right)^{\rm H}\! \left( \vect{W}_{b,u,k} -\vect{W}_{b,u,k}^t \right)\right) \\
    &\qquad- \lambda\left( \sum_{u=1}^U\sum_{k=1}^K\trace\left(\vect{W}_{b,u,k}^{\rm H}\vect{W}_{b,u,k} \right) - P_b^{\max} \right),
\end{align*}
where $\lambda$ is the nonnegative Lagrangian multiplier associated with the transmit power constraint. Then, the optimal $\vect{W}_{b,u,k}$ can be derived by comparing the partial derivative of $\mathcal{L}(\vect{W}_{b,u,k},\lambda)$ with respect to $\vect{W}_{b,u,k}$ against the zero matrix $\vect{0}_{N_t\times N_s}$, which results in the following closed-form expression for each BS precoding matrix:
\begin{equation} \label{eqn:optimal_precoder}
    \vect{W}_{b,u,k}^{\rm opt}(\lambda) = \left( \vect{E}_{b,u,k} + \lambda\vect{I}_{N_t} \right)^{-1} \vect{J}_{b,u,k},
\end{equation}
where the following matrix definitions have been used:
\begin{align}
    \vect{E}_{b,u,k} &\triangleq \frac{\rho^t}{\ln(2)} \widetilde{\vect{H}}_{b,u,k}^{\rm H}\vect{N}_{u,k}\widetilde{\vect{H}}_{b,u,k} + \frac{\tau}{2}\vect{I}_{N_t}, \label{eqn:optimal_precoder_part1} \\
    \vect{J}_{b,u,k} &\triangleq \frac{\rho^t}{\ln(2)} \widetilde{\vect{H}}_{b,u,k}^{\rm H}\left(\vect{M}_{u,k} - \vect{N}_{u,k}\vect{R}_{u,k}\right) \notag \\
    &\quad +\frac{\rho^t}{2}\vect{\Pi}_{\vect{W}_{b,u,k}}^t + \frac{1-\rho^t}{2}\vect{D}_{\vect{W}_{b,u,k}}^t + \frac{\tau}{2}\vect{W}_{b,u,k}^t. \label{eqn:optimal_precoder_part2}
\end{align}
Apparently, $\vect{W}_{b,u,k}^{\rm opt}$ depends on $\lambda$. To obtain its optimal value, it suffices to consider Slater's condition \cite{Boyd_2004} and then perform a bisection search similar to \cite{JSTSP_distributed} and references therein. 

\subsection{BDRIS Phase Configuration Optimization at Each $b$-th BS} \label{sec:RIS_Capacitors_Design}
The reduced optimization problem $\mathcal{OP}_{\mathcal{D},b}$ with respect to the capacitor matrix $\widetilde{\vect{C}}_b$, and eventually the passive beamforming matrix $\vect{\Phi}_{r,k}$, that needs to be solved by each $b$-th BS is expressed as follows:
\begin{align*}
	\mathcal{OP}_{\widetilde{\vect{C}}_b}: \,\max_{\widetilde{\vect{C}}_b} \, & \quad h(\widetilde{\vect{C}}_b) \triangleq \rho^t<\vect{\Gamma}_{\widetilde{\vect{C}}_b}^t + \vect{\Pi}_{\widetilde{\vect{C}}_b}^t,\widetilde{\vect{C}}_b-\widetilde{\vect{C}}_b^t> \\
    &+ (1-\rho^t)<\vect{D}_{\widetilde{\vect{C}}_b}^t,\widetilde{\vect{C}}_b-\widetilde{\vect{C}}_b^t> \\
    &- \frac{\tau}{2}\norm{\widetilde{\vect{C}}_b - \widetilde{\vect{C}}_b^t}_{\rm F}^2 \\
	\text{s.t.} & \quad \vect{C}_{b,r,g} = \vect{C}_{b,r,g}^{\rm T} \forall g\in\mathscr{G}, \\
	& \quad C_{\min} \leq [\vect{C}_{b,r,g}]_{n,n'} \leq C_{\max} \;\;\forall n,n'\in\mathscr{N}.
\end{align*}
It can be easily deduced that the above problem is concave with respect to $\widetilde{\vect{C}}_b$, since its objective function is concave and the constraint set is the intersection of two convex sets. In addition, to achieve consensus among the deployed BSs, we capitalize on first-order dynamic methods \cite{xin2020general} that allow tracking of the average gradient based on local information exchange. Specifically, as proposed in \cite{lorenzo2016next}, let the average gradient be denoted as $\overline{\nabla\mathcal{R}}\left(\widetilde{\vect{C}}_b^t;\vect{\eta}_b^t\right) \triangleq \frac{1}{B}\nabla_{\widetilde{\vect{C}}_b}\overline{\mathcal{R}}\left(\widetilde{\vect{C}}_b^t;\vect{\eta}_b^t\right)$, where $\overline{\mathcal{R}}\left(\widetilde{\vect{C}}_b;\vect{\eta}_b^t\right) \triangleq \sum_{u=1}^U\mathcal{R}_u\left(\widetilde{\vect{C}}_b;\vect{\eta}_b^t\right)$. Let also $\vect{Y}_{\widetilde{\vect{C}}_b}^t$ be a local auxiliary variable for each $b$-th BS, whose role is to asymptotically track $\overline{\nabla\mathcal{R}}\left(\widetilde{\vect{C}}_b^t;\vect{\eta}_b^t\right)$. Then, according to the dynamic average consensus method in \cite{zhu2010discrete}, gradient tracking can be enabled based on the following update rule:
\begin{equation} \label{eqn:Cap_Consensus}
\begin{aligned}
    \vect{Y}_{\widetilde{\vect{C}}_b}^t = &\sum_{i \in \mathcal{N}_b^t} [\vect{V}_{\rm net}^t]_{b,i}\vect{Y}_{\widetilde{\vect{C}}_i}^{t-1} + \nabla_{\widetilde{\vect{C}}_b}\overline{\mathcal{R}}\left(\widetilde{\vect{C}}_b^{t};\vect{\eta}_b^{t}\right) \\
    &- \nabla_{\widetilde{\vect{C}}_b}\overline{\mathcal{R}}\left(\widetilde{\vect{C}}_b^{t-1};\vect{\eta}_b^{t-1}\right)
\end{aligned}    
\end{equation}
with $\vect{Y}_{\widetilde{\vect{C}}_b}^{0} = \nabla_{\widetilde{\vect{C}}_b}\overline{\mathcal{R}}\left(\widetilde{\vect{C}}_b^{0};\vect{\eta}_b^{0}\right)$. In addition, the entries of the matrix $\vect{V}_{\rm net}^t \in \mathbb{R}^{B\times B}$ represent the weights (possibly time-varying) of the connected network graph $\mathcal{G}=(\mathscr{V},\mathscr{E})$, where $\mathscr{V}\equiv\mathscr{B}$ and $\mathscr{E}$ is the set of edges of the considered network topology, assuming that $\vect{V}_{\rm net}\vect{1}_B=\vect{1}_B$ and $\vect{1}_B^{\rm T}\vect{V}_{\rm net}^{\rm T} = \vect{1}_B^{\rm T}$, i.e., double stochasticity must hold for $\vect{V}_{\rm net}$ to enable consensus. In addition, $\mathcal{N}_b^t$ denotes the neighbors of each $b$-th BS. Thus, given $\vect{Y}_{\widetilde{\vect{C}}_b}^t$, each $b$-th BS can locally compute its pricing matrix $\vect{\Pi}_{\widetilde{\vect{C}}_b}^t$ as well as matrix $\vect{D}_{\widetilde{\vect{C}}_b}^t$ as follows:
\begin{align}
    \vect{\Pi}_{\widetilde{\vect{C}}_b}^t &= B\vect{Y}_{\widetilde{\vect{C}}_b}^t - \nabla_{\widetilde{\vect{C}}_b}\overline{\mathcal{R}}\left(\widetilde{\vect{C}}_b^{t};\vect{\eta}_b^t\right), \label{eqn:Pricing_Cap}\\
    \vect{D}_{\widetilde{\vect{C}}_b}^t &= (1-\rho^t)\vect{D}_{\widetilde{\vect{C}}_b}^{t-1} + \rho^tB\vect{Y}_{\widetilde{\vect{C}}_b}^{t}. \label{eqn:Accum_Cap}
\end{align} 
Clearly, to proceed with the solution of $\mathcal{OP}_{\widetilde{\vect{C}}_b}$, it suffices to derive the gradient of $\overline{\mathcal{R}}$ with respect to $\widetilde{\vect{C}}_b$, which is presented in the following theorem. 

\begin{Thm} \label{thm:gradient_Capacitors}
Let the following matrix definitions:
\begin{align}
    \vect{\Delta}_{b,r,q,k,g} &\triangleq \vect{W}_{b,q,k}^{\rm H} \vect{F}_{b,r,k,g}^{\rm H} \vect{Q}_{p,b,r,g}, \label{eqn:Delta_matrix}\\
    \vect{\Xi}_{b,r,u,k,g} &\triangleq \vect{G}_{r,u,k,g}^*\vect{Q}_{p,b,r,g}, \label{eqn:Xi_matrix}\\
    \vect{X}_{r,k,g} &\triangleq \vect{A}_{r,k,g} + \psi_0\vect{I}_{\tilde{M}}, \label{eqn:X_matrix}
\end{align}
where the following sub-matrices $\forall g \in \mathscr{G}$ have been used:
\begin{align}
    \vect{F}_{b,r,k,g} &\triangleq \left[\vect{F}_{b,r,k}\right]_{(1 + (g-1)\tilde{M}):g\tilde{M},:}, \label{eqn:submatrix_1} \\
    \vect{G}_{r,u,k,g} &\triangleq \left[\vect{G}_{r,u,k}\right]_{:,(1 + (g-1)\tilde{M}):g\tilde{M}}, \label{eqn:submatrix_2} \\
    \vect{Q}_{p,b,r,g} &\triangleq \left[\vect{Q}_{p,b,r} \right]_{(1 + (g-1)\tilde{M}):g\tilde{M},(1 + (g-1)\tilde{M}):g\tilde{M}}, \label{eqn:submatrix_3}
\end{align}
and $\vect{A}_{r,k,g}$ is given by \eqref{eqn:impedances}. Then, it holds that the gradient $\nabla_{\widetilde{\vect{C}}_b}\overline{\mathcal{R}}$ can be expressed as follows:
\begin{equation}
    \nabla_{\widetilde{\vect{C}}_b}\overline{\mathcal{R}} = \bigoplus_{r=1}^R \nabla_{\vect{C}_{b,r}}\overline{\mathcal{R}}
\end{equation}
with $\nabla_{\vect{C}_{b,r}}\overline{\mathcal{R}}\!=\! \bigoplus_{g=1}^G\nabla_{\vect{C}_{b,r,g}}\overline{\mathcal{R}}$ and $\nabla_{\vect{C}_{b,r,g}}\overline{\mathcal{R}}\!=\!\sum_{u=1}^U \sum_{k=1}^K \nabla_{\vect{C}_{b,r,g}}\mathcal{R}_{u,k}$, where each summand is given by:
\begin{equation} \label{eqn:Cap_gradient}
    \nabla_{\vect{C}_{b,r,g}}\mathcal{R}_{u,k} = \frac{-4\psi_0}{\ln(2)}\operatorname{unvec}\left(\Re\left\{ \vect{\Lambda}_{b,r,k,g}^{\rm H} \operatorname{vec}\left(\vect{\Theta}_{b,r,u,k,g} \right) \right\} \right).
\end{equation}
For the latter expression, it holds that: 
\begin{equation}
\begin{aligned} \label{eqn:Cap_gradient_part1}
    &\vect{\Theta}_{b,r,u,k,g} \triangleq \vect{X}_{r,k,g}^{-\rm H} \vect{\Xi}_{b,r,u,k,g}^{\rm T} \vect{P}_{u,k}^{-1}\vect{S}_{u,u,k}\vect{K}_{u,k}^{-1} \\
    &\times\left(\vect{\Delta}_{b,r,u,k,g} - \vect{S}_{u,u,k}^{\rm H}\vect{P}_{u,k}^{-1}\sum_{\substack{q=1,\\q\neq u}}^U\vect{S}_{u,q,k}\vect{\Delta}_{b,r,q,k,g} \right)\vect{X}_{r,k,g}^{-*},
\end{aligned}    
\end{equation}
while $\vect{\Lambda}_{b,r,k,g} \in \mathbb{C}^{\tilde{M}^2\times\tilde{M}^2}$ is a sparse matrix whose non-zero entries are the partial derivatives of $\vect{A}_{r,k,g}$ with respect to each entry of $\operatorname{vec}(\vect{C}_{b,r,g})$. In particular, it holds that:
\begin{equation} \label{eqn:Cap_gradient_part2}
[\vect{\Lambda}_{b,r,k,g}]_{i,j} =
\begin{cases}
\frac{\beta}{(\beta\delta[\vect{C}_{b,r,g}]_{n,n'}+1)^2}, &
\substack{i=(n'{-}1)\tilde{M}{+}n,\\ j=(\tilde{m}{-}1)\tilde{M}{+}n,\\ n=n',\ \forall\tilde{m}=1{:}\tilde{M}} \\[1ex]
\frac{-\beta}{(\beta\delta[\vect{C}_{b,r,g}]_{n,n'}+1)^2}, &
\substack{i=(n'{-}1)\tilde{M}{+}n,\\ j=(n'{-}1)\tilde{M}{+}n,\\ n\neq n'} \\[1ex]
0, & \text{otherwise}
\end{cases}
\end{equation}
where $\beta\triangleq \jmath\kappa f_k$, $\delta\triangleq R_0+\jmath\kappa f_k L_2$, and $i,j=1,2,\dots,\tilde{M}^2$.
\end{Thm}

\begin{IEEEproof}
    The proof is provided in Appendix~\ref{apx:gradient_Capacitors_proof}.
\end{IEEEproof}

Based on the previously derived gradient matrix $\vect{\Gamma}_{\widetilde{\vect{C}}_b}^t$, $\vect{\Pi}_{\widetilde{\vect{C}}_b}^t$ and $\vect{D}_{\widetilde{\vect{C}}_b}^t$ appearing in the objective function of $\mathcal{OP}_{\widetilde{\vect{C}}_b}$ can be trivially computed. Thus, it can be shown, after straightforward algebraic manipulations and usage of the identity $\trace(\vect{A}) = \trace(\vect{A}^{\rm T})$ as well as the condition $\widetilde{\vect{C}}_b^t = \widetilde{\vect{C}}_b^{t,\rm T}$, that $h(\widetilde{\vect{C}}_b) = -\frac{\tau}{2}\trace\left(\widetilde{\vect{C}}_b^{\rm T}\widetilde{\vect{C}}_b\right) + \trace\left(\vect{U}_{\widetilde{\vect{C}}_b}^{t,\rm T}\widetilde{\vect{C}}_b\right)$, where:
\begin{equation} \label{eqn:candidate_C}
    \vect{U}_{\widetilde{\vect{C}}_b}^t \triangleq \tau\widetilde{\vect{C}}_b^t + \rho^t\left(\vect{\Gamma}_{\widetilde{\vect{C}}_b}^t + \vect{\Pi}_{\widetilde{\vect{C}}_b}^t\right) + (1-\rho^t)\vect{D}_{\widetilde{\vect{C}}_b}^t.
\end{equation}

Next, according to the block diagonal structure of $\widetilde{\vect{C}}_b$, together with the facts that each $r$-th BDRIS and each $g$-th group are independent in the sense that they can be treated separately, $h(\widetilde{\vect{C}}_b)$ can be reduced to the expression:
\begin{equation}
\tilde{h}({\vect{C}_{b,r,g}}) = \sum_{g=1}^G \trace\left(\left(-\frac{\tau}{2}\vect{C}_{b,r,g} + \vect{U}_{\vect{C}_{b,r,g}}^t\right)^{\rm T} \vect{C}_{b,r,g}\right) ,
\end{equation}
where we used the properties: $(\bigoplus_{j=1}^J\vect{A}_j)^{\rm T} = \bigoplus_{j=1}^J \vect{A}_j^{\rm T}$ and $\trace(\bigoplus_{j=1}^J\vect{A}_j) = \sum_{j=1}^J\trace(\vect{A}_j)$ \cite{Zhang_2017}. As a result, it suffices to derive $\vect{C}_{b,r,g}^{\rm opt}$ $\forall r\in\mathscr{R}$, $\forall g\in\mathscr{G}$ and then form $\widetilde{\vect{C}}_b^{\rm opt}$. To this end, a candidate solution is to consider the unconstrained solution, using the first-order condition of optimality which yields $\vect{C}_{b,r,g}^{\star} = \frac{1}{\tau}\vect{U}_{\vect{C}_{b,r,g}}^t$. However, we note that the matrix $\vect{U}_{\vect{C}_{b,r,g}}^t$ on the right-hand side is not feasible. Even if it satisfies the box constraints, it does not satisfy the symmetry constraint (i.e., $\vect{C}_{b,r,g}^{\star}\neq \vect{C}_{b,r,g}^{\star,\rm T}$), which is evident after observing \eqref{eqn:Cap_gradient}--\eqref{eqn:Cap_gradient_part2}. Even if the constraint set for the design matrix $\vect{C}_{b,r,g}$ is convex (as explained earlier), it is not trivial to project $\vect{C}_{b,r,g}^{\star}$, because there does not exist any delicate closed-form expression. To address this difficulty and, at the same time, guarantee a strongly convergent point $\vect{C}_{b,r,g}^{\rm opt}$ \cite{bauschke2011convex}, we propose an iterative alternating projection procedure, according to the reduced version of Dykstra's algorithm \cite{dattorro2010convex}, whose steps are summarized in Algorithm~\ref{alg:Dykstras}. Therein, in Steps $3$ and $5$, we use the mappings $\mathcal{P}_{\mathcal{S}}(\cdot)$ and $\mathcal{P}_{\mathcal{B}}(\cdot)$ which act as projection operators on the symmetric and box constraints' sets for their matrix inputs, respectively. In particular, $\mathcal{P}_{\mathcal{S}}(\vect{A}) = \frac{1}{2}(\vect{A} + \vect{A}^{\rm T})$, while $\mathcal{P}_{\mathcal{B}}(\vect{A}) = \max(\min([\vect{A}]_{i,j}, C_{\max}), C_{\min})$ $\forall i,j = 1,2,\dots,N$, for an arbitrary square matrix $\vect{A}\in\mathbb{R}^{N\times N}$. Therefore, the output of Algorithm~\ref{alg:Dykstras} is the desired feasible solution to $\mathcal{OP}_{\widetilde{\vect{C}}_b}$. 
%%%%%%%%%%% ALGORITHM %%%%%%%%%%%
\begin{algorithm}[!t]
	\begin{algorithmic}[1]
		\caption{Proposed Dykstra's-based Solution for $\vect{C}_{b,r,g}$}
		\label{alg:Dykstras}
		\State \textbf{Input:} $\ell = 0$, $\vect{C}_{b,r,g}^{(0)} = \vect{C}_{b,r,g}^{\star}$, $\vect{V}^{(0)}\!=\!\vect{B}^{(0)}\!=\!\vect{0}_{\tilde{M}\times\tilde{M}}$, and algorithmic threshold $\epsilon > 0$.
		\For{$ \ell = 0, 1, \dots$}
        \State Compute $\vect{A}^{(\ell)} = \mathcal{P}_{\mathcal{S}}\left(\vect{C}_{b,r,g}^{(\ell)} + \vect{V}^{(\ell)}\right)$.
        \State Compute $\vect{V}^{(\ell +1)} = \vect{C}_{b,r,g}^{(\ell)} + \vect{V}^{(\ell)} - \vect{A}^{(\ell)}$.
        \State Compute $\vect{C}_{b,r,g}^{(\ell +1)} = \mathcal{P}_{\mathcal{B}}\left( \vect{A}^{(\ell)} + \vect{B}^{(\ell)}\right)$.
        \State Compute $\vect{B}^{(\ell + 1)} = \vect{A}^{(\ell)} + \vect{B}^{(\ell)} - \vect{C}_{b,r,g}^{(\ell + 1)}$.
        \If $\norm{\vect{C}_{b,r,g}^{(\ell+1)} - \vect{C}_{b,r,g}^{(\ell)}}_{\rm F} \leq \epsilon$
		\State $\vect{C}_{b,r,g}^{\rm opt} = \vect{C}_{b,r,g}^{(\ell+1)}$ and \textbf{break};
		\EndIf
		\EndFor
		\State \textbf{Output:} $\vect{C}_{b,r,g}^{\rm opt}$.
	\end{algorithmic}
\end{algorithm}
%%%%%%%%%%% END OF ALGORITHM %%%%%%%%%%%

\subsection{RIS Permutation Matrix Optimization at Each $b$-th BS} \label{sec:RIS_Permutation_Design}
The design problem including the permutation matrices $\vect{Q}_{p,b,r}$ $\forall r\in\mathscr{R}$ included in $\widetilde{\vect{Q}}_{p,b}$ can be shown (similarly to the case of $\mathcal{OP}_{\widetilde{\vect{C}}_b}$) that it reduces to the following simplified optimization sub-problem:
\begin{align*}
    \mathcal{OP}_{\widetilde{\vect{Q}}_{p,b}}: \,\max_{\vect{Q}_{p,b,r}\in\mathcal{Q}} \, & \quad g(\vect{Q}_{p,b,r}) \triangleq \trace\left(\vect{U}_{\vect{Q}_{p,b,r}}^{t,\rm T}\vect{Q}_{p,b,r}\right)\\
    &- \frac{\tau}{2}\norm{\vect{Q}_{p,b,r} - \vect{Q}_{p,b,r}^t}_{\rm F}^2,
\end{align*}
where $\vect{U}_{\vect{Q}_{p,b,r}}^t \triangleq \rho^t\left(\vect{\Gamma}_{\vect{Q}_{p,b,r}}^t + \vect{\Pi}_{\vect{Q}_{p,b,r}}^t \right) + (1-\rho^t)\vect{D}_{\vect{Q}_{p,b,r}}^t$. To obtain the solution of $\mathcal{OP}_{\widetilde{\vect{Q}}_{p,b}}$, we first need to derive the matrices $\vect{\Gamma}_{\vect{Q}_{p,b,r}}^t$, $\vect{\Pi}_{\vect{Q}_{p,b,r}}^t$, and $\vect{D}_{\vect{Q}_{p,b,r}}^t$, whose expressions depend on the gradient $\nabla_{\vect{Q}_{p,b,r}}\mathcal{R}_{u,k}$, as derived in the sequel. 

\begin{Cor} \label{cor:Perm_Gradient}
The gradient of $\mathcal{R}_{u,k}$ with respect to the matrix $\vect{Q}_{p,b,r}$ is given by the following analytical expression:
\begin{equation}
\begin{aligned}
    &\nabla_{\vect{Q}_{p,b,r}}\mathcal{R}_{u,k} = \frac{2}{\ln(2)}\Re\Bigg\{\vect{\Phi}_{r,k}\vect{Q}_{p,b,r}^{\rm T}\\
    &\times\left(\vect{\Sigma}_{b,r,u,k} + \vect{\Sigma}_{b,r,u,k}^{\rm T} - \sum_{\substack{q=1,\\q\neq u}}^U \left(\vect{\Psi}_{b,r,u,q,k} + \vect{\Psi}_{b,r,u,q,k}^{\rm T}\right) \right)\Bigg\},
\end{aligned}
\end{equation}
where the following matrix definitions have been used:
\begin{align}
    &\,\,\,\,\vect{\Sigma}_{b,r,u,k} \triangleq \vect{F}_{b,r,k}\vect{W}_{b,u,k}\vect{K}_{u,k}^{-1}\vect{S}_{u,u,k}^{\rm H}\vect{P}_{u,k}^{-1}\vect{G}_{r,u,k}, \label{eqn:Sigma} \\
    &\begin{aligned}
        \vect{\Psi}_{b,r,u,q,k} &\triangleq \vect{F}_{b,r,k}\vect{W}_{b,q,k}\vect{S}_{u,q,k}^{\rm H}\vect{P}_{u,k}^{-1}\\
        &\,\,\times\vect{S}_{u,u,k}\vect{K}_{u,k}^{-1}\vect{S}_{u,u,k}^{\rm H}\vect{P}_{u,k}^{-1}\vect{G}_{r,u,k}. \label{eqn:Psi}
    \end{aligned}
\end{align}
\end{Cor}

\begin{IEEEproof}
    The detailed proof is omitted due to space limitations, since it is similar to that of Theorem~\ref{thm:gradient_Capacitors} using the expression \eqref{eqn:channel_unfolded} in Appendix~\ref{apx:gradient_Capacitors_proof} for the channel $\widetilde{\vect{H}}_{b,u,k}$. 
\end{IEEEproof}

Based on the above derivation for the gradient, we can compute the expressions for the matrices included in the matrix $\vect{U}_{\vect{Q}_{p,b,r}}^t$ that appears in the objective function $g(\vect{Q}_{p,b,r})$, as:
\begin{align}
    \vect{\Gamma}_{\vect{Q}_{p,b,r}}^t &= \sum_{u=1}^U\sum_{k=1}^K \nabla_{\vect{Q}_{p,b,r}}\mathcal{R}_{u,k}\left(\vect{Q}_{p,b,r}^t;\vect{\eta}_b^t\right), \label{eqn:Gamma_Permut} \\
    \vect{\Pi}_{\vect{Q}_{p,b,r}}^t &= B\vect{Y}_{\vect{Q}_{p,b,r}}^t - \vect{\Gamma}_{\vect{Q}_{p,b,r}}^t, \label{eqn:Pricing_Permut} \\
    \vect{D}_{\vect{Q}_{p,b,r}}^t &= (1-\rho^t)\vect{D}_{\vect{Q}_{p,b,r}}^{t-1} + \rho^tB\vect{Y}_{\vect{Q}_{p,b,r}}^t, \label{eqn:Accum_Permut}
\end{align}
where $\vect{Y}_{\vect{Q}_{p,b,r}}^t$ is expressed, similar to \eqref{eqn:Cap_Consensus}, as follows:
\begin{equation} \label{eqn:Perm_Consensus}
   \vect{Y}_{\vect{Q}_{p,b,r}}^t = \sum_{i\in\mathcal{N}_b^t}[\vect{V}_{\rm net}]_{b,i}\vect{Y}_{\vect{Q}_{p,i,r}}^{t-1} + \vect{\Gamma}_{\vect{Q}_{p,b,r}}^t - \vect{\Gamma}_{\vect{Q}_{p,b,r}}^{t-1}.
\end{equation}

%%%%%%%%%%% ALGORITHM %%%%%%%%%%%
\begin{algorithm}[!t]
	\begin{algorithmic}[1]
		\caption{Proposed CSD-SCA Design Solving $\mathcal{OP}$}
		\label{alg:OP_Overall_Distr_Alg}
		\State \textbf{Input:} $t=0$, $\{\alpha^t\}> 0$, $\{\rho^t\}> 0$, $\tau>0$, $\epsilon > 0$, $B$, feasible $\widetilde{\vect{W}}^0$, $\widetilde{\vect{C}}^0$, $\widetilde{\vect{Q}}_p^0$, $\overline{\mathcal{R}}^0=\sum_{u=1}^U\mathcal{R}_u(\widetilde{\vect{W}}^0,\widetilde{\vect{C}}^0,\widetilde{\vect{Q}}_p^0;\vect{\eta}_b^0)$, $\vect{Y}_{\widetilde{\vect{C}}_b}^0=\nabla_{\widetilde{\vect{C}}_b}\overline{\mathcal{R}}(\widetilde{\vect{C}}_b^0;\vect{\eta}_b^0)$, $\vect{\Pi}_{\widetilde{\vect{C}}_b}^0=(B-1)\vect{Y}_{\widetilde{\vect{C}}_b}^0$, and $\vect{D}_{\widetilde{\vect{C}}_b}^0 = B\vect{Y}_{\widetilde{\vect{C}}_b}^0$, as well as $\vect{Y}_{\widetilde{\vect{Q}}_{p,b}}^0=\nabla_{\widetilde{\vect{Q}}_{p,b}}\overline{\mathcal{R}}(\widetilde{\vect{Q}}_{p,b}^0;\vect{\eta}_b^0)$, $\vect{\Pi}_{\widetilde{\vect{Q}}_{p,b}}^0=(B-1)\vect{Y}_{\widetilde{\vect{Q}}_{p,b}}^0$ and $\vect{D}_{\widetilde{\vect{Q}}_{p,b}}^0 = B\vect{Y}_{\widetilde{\vect{Q}}_{p,b}}^0$ $\forall b\in \mathscr{B}$.
		\For{$ t = 0,1,\ldots$}
            \State 
            \parbox[t]{\dimexpr\linewidth-\algorithmicindent}{%
            Compute $\widetilde{\vect{\Phi}}_{r,k}^t(\widetilde{\vect{C}}_b^t,\widetilde{\vect{Q}}_{p,b}^t)\,\,\forall r,k,b$ using \eqref{eqn:Phi_complete}, \eqref{eqn:RIS_Scattering}, and \eqref{eqn:impedances}.
            }
            \State \textbf{(S.1) SCA Optimization:}
		\For{$ b = 1,2,\ldots,B$}
            \State 
            \parbox[t]{\dimexpr\linewidth-\algorithmicindent}{%
            Compute $\vect{\Pi}_{\vect{W}_{b,u,k}}^t$ according to \eqref{eqn:pricing_precoder} and\\ $\vect{D}_{\vect{W}_{b,u,k}}^t$ combining \eqref{eqn:pricing_precoder} and \eqref{eqn:gradient_W_Accum}.
            }
            \State 
            \parbox[t]{\dimexpr\linewidth-\algorithmicindent}{%
            Compute $\hat{\vect{W}}_{b,u,k}^{t}=\arg\max\limits_{\vect{W}_{b,u,k}} f(\vect{W}_{b,u,k})\forall u,k$,\\ according to \eqref{eqn:optimal_precoder} and a bisection method.
            }
        \State
        \parbox[t]{\dimexpr\linewidth-\algorithmicindent}{%
        Compute $\vect{\Gamma}_{\widetilde{\vect{C}}_b}^t$ according to Theorem~\ref{thm:gradient_Capacitors}.
        }
		\State 
            \parbox[t]{\dimexpr\linewidth-\algorithmicindent}{%
            Obtain $\hat{\vect{C}}_b^t = \arg\max\limits_{\widetilde{\vect{C}}_b} h(\widetilde{\vect{C}}_b)$ according to \\ Algorithm~\ref{alg:Dykstras} using as input $\frac{1}{\tau}\vect{U}_{\widetilde{\vect{C}}_b}^t$ given in \eqref{eqn:candidate_C}.
            }
        \State Compute $\vect{\Gamma}_{\vect{Q}_{p,b}}^t$ according to \eqref{eqn:Gamma_Permut}.
        \State 
            \parbox[t]{\dimexpr\linewidth-\algorithmicindent}{%
            Compute $\hat{\vect{Q}}_{p,b,r}^t=\arg\max\limits_{{\vect{Q}_{p,b,r}}} g(\vect{Q}_{p,b,r})$ solving \\the LSAP problem $\mathcal{OP}_{\widetilde{\vect{Q}}_{p,b}}$ numerically.
            }
		\State 
            \parbox[t]{\dimexpr\linewidth-\algorithmicindent}{%
            Obtain $\vect{W}_{b,u,k}^{t+1} = (1-\alpha^t)\vect{W}_{b,u,k}^t\!+\!\alpha^t \hat{\vect{W}}_{b,u,k}^{t}$,\\ $\breve{\vect{C}}_{b}^{t+1} = (1-\alpha^t)\widetilde{\vect{C}}_{b}^t + \alpha^t\hat{\vect{C}}_{b}^t$, and\\ $\breve{\vect{Q}}_{p,b}^{t+1} = (1-\alpha^t)\widetilde{\vect{Q}}_{p,b}^t + \alpha^t\hat{\vect{Q}}_{p,b}^t$.
            }
		\EndFor
        \State The random set of channels $\vect{\eta}^{t+1}$ is realized.
        \State \textbf{(S.2) Consensus Updates for the $R$ BDRISs:}
        \For{$ b = 1,2,\ldots,B$}
            \State Obtain $\widetilde{\vect{C}}_b^{t+1} = \sum_{i\in\mathcal{N}_b^t} [\vect{V}_{\rm net}^t]_{b,i} \breve{\vect{C}}_i^{t+1}$.
            \State 
                \parbox[t]{\dimexpr\linewidth-\algorithmicindent}{%
                Compute $\vect{Y}_{\widetilde{\vect{C}}_b}^{t+1}$ and $\vect{\Pi}_{\widetilde{\vect{C}}_b}^{t+1}$ according to \\\eqref{eqn:Cap_Consensus} and \eqref{eqn:Pricing_Cap}, respectively.
                }
            \State Compute $\check{\vect{Q}}_{p,b}^{t+1}=\sum_{i\in\mathcal{N}_b^t}[\vect{V}_{\rm net}^t]_{b,i}\breve{\vect{Q}}_{p,i}^{t+1}$.
            \State
            \parbox[t]{\dimexpr\linewidth-\algorithmicindent}{%
            Obtain $\widetilde{\vect{Q}}_{p,b}^{t+1}$ solving the LSAP problem\\$\mathcal{OP}_{\vect{Q}_{p,b}}^{\rm Proj}$ numerically.
            }
            \State 
            \parbox[t]{\dimexpr\linewidth-\algorithmicindent}{%
            Compute $\vect{\Pi}_{\vect{Q}_{p,b,r}}^{t+1}$ and $\vect{Y}_{\vect{Q}_{p,b,r}}^{t+1}\forall r \in \mathscr{R}$ using \\\eqref{eqn:Pricing_Permut} and \eqref{eqn:Perm_Consensus}, respectively.
            }
        \EndFor
        \State \textbf{(S.3) Gradient Averaging:}
        \For{$ b = 1,2,\ldots,B$}
            \State 
            \parbox[t]{\dimexpr\linewidth-\algorithmicindent}{%
            Compute $\vect{D}_{\vect{W}_{b,u,k}}^{t+1}\forall u,k$ according to \eqref{eqn:accum_definition}.
            }
            \State 
            \parbox[t]{\dimexpr\linewidth-\algorithmicindent}{%
            Compute $\vect{D}_{\widetilde{\vect{C}}_b}^{t+1}$ and $\vect{D}_{\vect{Q}_{p,b,r}}^{t+1}\forall r\in\mathscr{R}$ according \\to \eqref{eqn:Accum_Cap} and \eqref{eqn:Accum_Permut}, respectively.
            }
        \EndFor
		\If $\left\lvert\left(\overline{\mathcal{R}}^{(t+1)} - \overline{\mathcal{R}}^{(t)}\right)/\overline{\mathcal{R}}^{(t+1)}\right\rvert \leq \epsilon$, \textbf{break}; 
		\EndIf
		\EndFor
		\State \textbf{Output:} $\vect{X}^{t^{\star}}=\left[\widetilde{\vect{W}}^{t^{\star},\rm T},\widetilde{\vect{C}}^{t^{\star},\rm T},\widetilde{\vect{Q}}_p^{t^{\star},\rm T}\right]^{\rm T}$.
	\end{algorithmic}
\end{algorithm}
%%%%%%%%%%% END OF ALGORITHM %%%%%%%%%%%

Therefore, to solve $\mathcal{OP}_{\widetilde{\vect{Q}}_{p,b}}$, it suffices to expand the norm, which indicates that this problem can be equivalently rewritten as follows: 
\begin{align*}
	\mathcal{OP}_{\widetilde{\vect{Q}}_{p,b}}: \,\max_{\vect{Q}_{p,b,r} \in \mathcal{Q}} \trace\left(\widetilde{\vect{U}}_{\vect{Q}_{p,b,r}}^{t,\rm T} \vect{Q}_{p,b,r}\right)
\end{align*}
with $\widetilde{\vect{U}}_{\vect{Q}_{p,b,r}}^t \triangleq \tau\vect{Q}_{p,b,r}^t + \vect{U}_{\vect{Q}_{p,b,r}}^t$, where the quadratic term $\trace(\vect{Q}_{p,b,r}^{\rm T}\vect{Q}_{p,b,r})$ has been treated as constant equal to $M$, since, by definition, $\vect{Q}_{p,b,r}$ is a permutation matrix. This simplified version of $\mathcal{OP}_{\widetilde{\vect{Q}}_{p,b}}$ belongs to the class of Linear Assignment Problems (LSAPs) and its optimal solution can be efficiently obtained either by linear programming techniques or permutations of sparse matrix strategies \cite{duff2001algorithms}. 

\subsection{Overall Decentralized Cooperative Solution} \label{seq:Overall_Solution}
According to the previous subsections, the best response mapping $\hat{\vect{X}}_b^t$ can be obtained by solving $\mathcal{OP}_{\tilde{\vect{W}}_b}$, $\mathcal{OP}_{\widetilde{\vect{C}}_b}$, and $\mathcal{OP}_{\widetilde{\vect{Q}}_{p,b}}$. Then, the solution to $\mathcal{OP}$ can be finalized by applying the update rule $(1-\alpha^t)\vect{X}_b^t + \alpha^t\hat{\vect{X}}_b^t$ to each variable, where $\alpha^t$ represents the step size varying over algorithmic iterations. However, for the design parameters $\widetilde{\vect{C}}$ and $\widetilde{\vect{Q}}$ of the BDRISs, consensus updates are also needed to force asymptotic agreement among the BSs. Specifically, each BS updates its BDRIS variables according to the consensus-based step expressed as $\vect{X}_b^{t+1} = \sum_{i\in\mathcal{N}_b^t} [\vect{V}_{\rm net}^t]_{b,i}\breve{\vect{X}}_i^{t+1}$, where $\vect{X}_b$ is either the capacitors or the permutation matrix design parameter, and $\breve{\vect{X}}_b$ represents the resulting point after the update rule with step size $\alpha^t$. In addition, $\mathcal{N}_b^t \triangleq \{i|(i,b)\in\mathscr{E}^t\}\cup\{b\}$ represents the neighborhood of each $b$-th BS at the algorithmic step $t$ that also includes this BS. At this point, it is noted that the last two updating rules for the BDRIS variable $\widetilde{\vect{Q}}_{p,b}$ violate the feasibility of this design parameter, i.e., it most likely holds $\widetilde{\vect{Q}}_{p,b}^{t+1} \notin \mathcal{Q}$. To address this issue and obtain a feasible $\widetilde{\vect{Q}}_{p,b}^{t+1}$ that satisfies consensus, we further solve the following projection problem:
\begin{align*}
    \mathcal{OP}_{\vect{Q}_{p,b,r}}^{\rm Proj}: \,\widetilde{\vect{Q}}_{p,b,r}^{t+1} = \arg\min_{\vect{Q}_{p,b,r}\in\mathcal{Q}} \norm{\vect{Q}_{p,b,r} - \check{\vect{Q}}_{p,b,r}^{t+1}}_{\rm F}^2,
\end{align*}
where $\check{\vect{Q}}_{p,b,r}^{t+1}$ denotes the infeasible variable obtained after the update rules mentioned above. It can be shown that $\mathcal{OP}_{\vect{Q}_{p,b,r}}^{\rm Proj}$ reduces to the class of LSAPs, which can be efficiently solved similar to $\mathcal{OP}_{\widetilde{\vect{Q}}_{p,b}}$. The overall distributed cooperative solution of $\mathcal{OP}$, termed as Consensus-based Stochastic Distributed Successive Concave Approximation (CSD-SCA), is summarized in Algorithm~\ref{alg:OP_Overall_Distr_Alg}.

\subsection{Convergence and Complexity Analysis} \label{sec:conv_comp_analysis}
Let us first define the output sequence $\vect{x}^{t^{\star}} \triangleq \operatorname{vec}\left(\left\{\vect{X}^{t^{\star}}\right\}\right)$ of Algorithm~\ref{alg:OP_Overall_Distr_Alg}, and let $\vect{x}_{\rm RIS}^{t^{\star}}\in\mathbb{R}^{Bd\times 1}$ be the vector that collects all output BDRIS variables, i.e., $\widetilde{\vect{C}}^{t^{\star}}$ and $\widetilde{\vect{Q}}_p^{t^{\star}}$, where $t^{\star}$ denotes the number of algorithmic iterations that Algorithm~\ref{alg:OP_Overall_Distr_Alg} terminates and $d$ is the dimension of the concatenated vector of the BDRIS variables at each $b$-th BS. Let also $\bar{\vect{x}}_{\rm RIS}^{t^{\star}}\triangleq \frac{1}{B}(\vect{1}_B^{\rm T}\otimes\vect{I}_d)\vect{x}_{\rm RIS}^{t^{\star}} \in\mathbb{R}^{d\times 1}$ be the average vector of the BDRIS parameters. Then, we have the following theorem.
\begin{Thm} \label{thm:conv_consensus}
    Assume that: \textit{i}) the channel realizations $\{\vect{\eta}^t\}_{t=0}^{t^{\star}}$ are bounded, independent, and identically distributed; \textit{ii}) the graph $\mathcal{G}$ is connected; and \textit{iii}) the weight matrix $\vect{V}_{\rm net}^t$ satisfies the double stochasticity property: $\vect{V}_{\rm net}^t\vect{1}_B=\vect{V}_{\rm net}^{t,\rm T}\vect{1}_B=\vect{1}_B$. Moreover, if the sequences $\{\alpha^t\}$, $\{\rho^t\}$ are chosen so that the following conditions are satisfied: $\gamma^t\in(0,1]$~$\forall t$, $\sum_{t=0}^{\infty}\gamma^t<\infty$, $\sum_{t=0}^{\infty}(\gamma^t)^2<\infty$ (where $\gamma^t$ is equal to $\alpha^t$ or $\rho^t$), and $\lim_{t\rightarrow\infty} \alpha^t/\rho^t=0$, then the following two properties hold for the proposed CSD-SCA algorithm:
    \begin{enumerate}
        \item Convergence: every limit point of the sequence $\{\vect{x}^{t^{\star}}\}$ is a stationary solution of $\mathcal{OP}$ almost surely.
        \item Consensus: the consensus error between the sequences $\{\vect{x}_{\rm RIS}^{t^{\star}}\}$ and $\{\bar{\vect{x}}_{\rm RIS}^{t^{\star}}\}$ asymptotically vanishes, i.e., $\norm{\vect{x}_{\rm RIS}^{t^{\star}} - (\vect{1}_B\otimes\vect{I}_d)\bar{\vect{x}}_{\rm RIS}^{t^{\star}}} \xrightarrow[t^{\star}\rightarrow\infty]{}0$.
    \end{enumerate}
\end{Thm}

\begin{IEEEproof}
    The proof follows as a combination of the proofs included in \cite[Theorem 2]{JSTSP_distributed}, \cite[Theorem 1]{yang2016parallel}, and \cite[Theorem 3]{lorenzo2016next}, and thus omitted due to space limitations.
\end{IEEEproof}

The computational complexity of Algorithm~\ref{alg:OP_Overall_Distr_Alg} is analyzed, based on its main steps, as follows. In Step $3$, the computation of the cascaded channels involves the inversion of the admittance matrices, which requires $\mathcal{O}(BRKG\tilde{M}^3)$ computations. The computations of the pricing and accumulation matrices related to Step $6$ results in $\mathcal{O}(2UKN_r^3)$ complexity, while Step $7$ requires $\mathcal{O}(UKN_t^3)$ computational cost. In Step $8$, the worst case complexity is $\mathcal{O}(RG\tilde{M}^2)$ due to the multiplication in \eqref{eqn:Cap_gradient}, while the computational cost of Algorithm~\ref{alg:Dykstras} invoked in Step $9$ is negligible. In addition, Step $10$ requires the worst case complexity of $\mathcal{OP}(2N_s^3 N_r^3)$ due to the matrix inversions in \eqref{eqn:Sigma} and \eqref{eqn:Psi}, while the solution to the LSAP problem $\mathcal{OP}_{\widetilde{\vect{Q}}_{p,b}}$ has worst case complexity $\mathcal{O}(RM(\mu + M\log M))$ according to \cite{JSTSP_distributed} and references therein. As a result, the overall computational complexity for the SCA optimization algorithmic steps (i.e., Step $5$--$13$) is equal to $\mathcal{O}(B(2UKN_r^3 + UKN_t^3 + RG\tilde{M}^2 + 2N_s^3N_r^3 + RM(\mu + MlogM)))$. Similarly, it can be deduced that the consensus updates (Step $16$--$22$) and gradient averaging (Step $24$--$27$) have computational complexity equal to $\mathcal{O}(B(N_s^3N_r^3 + RM(\mu + \log M)))$ and $\mathcal{O}(BN_s^3N_r^3)$, respectively. Putting all above together, the overall complexity of the proposed distributed cooperative design solving $\mathcal{OP}$ is characterized by the following expression:
\begin{align}
    &{\rm CC}_{\mathcal{OP}} = \mathcal{O}\left(T_{\max}B\left(RKG\tilde{M}^3 + 2UKN_r^3 + UKN_t^3\right.\right.\nonumber\\ 
    &\left.\left.+ RG\tilde{M}^2 + 4N_s^3N_r^3 + 2RM(\mu +M\log M)\right)\right),
\end{align}
where $T_{\max}$ represents the total number of algorithmic iterations until convergence. 

\subsection{Adaptive Weights for $\vect{V}_{\rm net}^t$} \label{sec:V_net_Adaptive}
The selection of the matrix $\vect{V}_{\rm net}^t$ plays a crucial role in the consensus behavior of Algorithm~\ref{alg:OP_Overall_Distr_Alg}, and particularly, in the evolution of the consensus error as $t$ increases. In this paper, we propose a heuristic, yet efficient method for the values of $\vect{V}_{\rm net}^t$ that can be applied in each algorithmic iteration $t$; the method takes into account the new channel realization $\vect{\eta}^{t+1}$ in Step $14$ of Algorithm~\ref{alg:OP_Overall_Distr_Alg}. In particular, it suffices to first build the adjacency matrix of the digraph $\mathcal{G}$; then, $\vect{V}_{\rm net}^t$ can be obtained according to the Metropolis-Hastings or uniform weights (see \cite{lorenzo2016next} and references therein). To obtain the adjacency matrix (whose dimensions are $B\times B$), we initially form all cascaded channels $\widetilde{\vect{H}}_{b,u,k}$ and average them over all $K$ SCs to build the matrix $\widehat{\vect{H}}_{b,u}$. Then, under the reasonable assumption that $U\geq B$, we randomly select $B$ indices of $\widehat{\vect{H}}_{b,u}$ among $U$ of them and measure the channel gains $\|\widehat{\vect{H}}_{b,\bar{b}}\|_{\rm F}$ with $\bar{b} \in \mathscr{B}$, which are stored in the matrix $\vect{\Omega}_{\widehat{\vect{H}}}\in\mathbb{R}^{B\times B}$. Next, we compute the threshold value $\mathcal{H}_{\rm thr}\triangleq \frac{1}{2}(\max(\vect{\Omega}_{\widehat{\vect{H}}}) - \min(\vect{\Omega}_{\widehat{\vect{H}}}))$ allowing us to calculate the adjacency via the following process: if $[\vect{\Omega}_{\widehat{\vect{H}}}]_{b,\bar{b}}\geq\mathcal{H}_{\rm thr}$, then the corresponding entry of the adjacency matrix is set equal to one; otherwise it remains zero.
\begin{figure}[!t]
	\centering
	\includegraphics[trim=0.5cm 0.9cm 0.5cm 1.25cm,clip,width=3.75in]{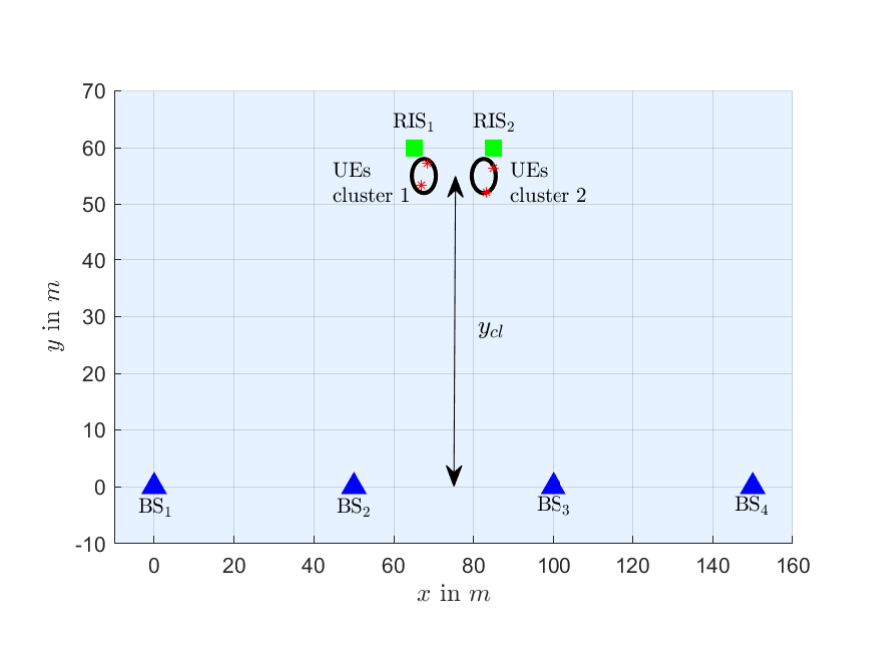}
	\caption{\small{The $xy$-plane of the simulated in $3$D cell-free MIMO system including $B=4$ BSs and $R=2$ shared BDRISs. Each node's coordinates include the distances $x$, $y$, and $z$ along the corresponding axes. The positions of the UEs have been randomly selected from two clusters, each with $U=4$ UEs, with the setting $y_{cl}=57.5\,m$.}}
	\label{fig:Simulation_Setup}
\end{figure}

\section{Numerical Results and Discussion} \label{sec:Numerical}
In this section, we present sum-rate performance evaluation results for the proposed robust decentralized cooperative design presented in Section~\ref{sec:Distr_OP_Solution}. We have particularly simulated, for various scenarios, Algorithm~\ref{alg:OP_Overall_Distr_Alg} to design the precoding matrices of the multiple BSs of the considered wideband cell-free MIMO communication system as well as the tunable capacitances and permutation matrices of the multiple BDRISs.

\subsection{Simulation Setup} \label{sec:Sims_Setup}
A $3$-Dimensional ($3$D) Cartesian coordinate system was considered for the placement of the system nodes, whose $xy$-plane is illustrated in Fig.~\ref{fig:Simulation_Setup}. The coordinates of each node are thus given by the triad $(x,y,z)$, with each entry representing the location on the corresponding axis. As depicted in the figure, we assumed the presence of $B=4$ BSs with each $b$-th BS located in $(50(b-1)\,m,0\,m,5\,m)$ and equipped with $N_t=2$ antennas, while serving $U=4$ UEs each possessing $N_r=2$ antenna elements. The UEs were divided and randomly placed into two separate circular groups, each with the same radius equal to $2\,m$. These two circular clusters were located at $y_{cl}=57.5\,m$ along the $y$-axis from the line of the BSs' placement, as shown in Fig.~\ref{fig:Simulation_Setup}, while their $x$-coordinates were set at $67.5\,m$ for the left one and $82.5\,m$ for the second one. The cell-free system also included $R=2$ BDRISs whose coordinates were fixed at $(65,60,6)\,m$ for $\rm{BDRIS_1}$ and $(85,60,6)\,m$ for $\rm{BDRIS_2}$, both having the same number $M=64$ of unit elements and $G=4$. All wireless links were modeled as wideband Rayleigh fading channels with distance-dependent pathloss, which was modeled as $\operatorname{PL}_{i,j} = \operatorname{PL}_0(d_{i,j}/d_0)^{-\alpha_{i,j}}$ with $\operatorname{PL}_0=-30\,{\rm dB}$, $d_{i,j}$ being the distance between any two nodes $i,j$ (with $(i,j)\in\{\rm{BS},\rm{UE},\rm{RIS}\}$), and $d_0 = 1\,m$. For the pathloss exponents, we have set $\alpha_{\rm{BS,UE}} = 3.8$, $\alpha_{\rm{BS,RIS}} = 2.4$, and $\alpha_{\rm{RIS,UE}} = 2.2$, while to produce noisy channel samples at each $b$-th BS (i.e., $\vect{\eta}_b^t)$ for our robust framework, we have used the model described in \cite[eq. (53)]{zhang2021joint} according to which: $\hat{g} = g + e$ with $g$ being the actually estimated channel gain and $e\sim\mathcal{CN}(0,\sigma_e^2)$, where $\sigma_e^2\triangleq\delta\abs{g}^2$ and $\delta=0.2$. 
%while to produce imperfect channel samples (i.e., $\vect{\eta}^t)$, we have used the model described in \cite[eq. (53)]{zhang2021joint} according to which: $\hat{g} = g + e$ with $g$ being the actual channel and $e\sim\mathcal{CN}(0,\sigma_e^2)$, where $\sigma_e^2\triangleq\delta\abs{g}^2$ and $\delta=0.2$. 
\begin{figure}[!t]
	\centering
	\includegraphics[trim=0cm 0cm 0cm 0.5cm,clip,width=3.75in]{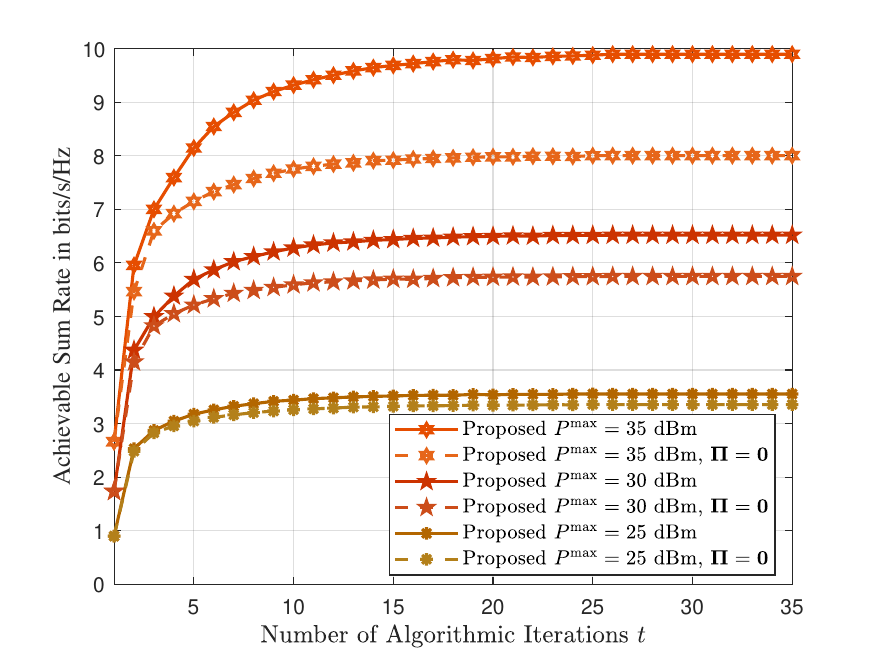}
	\caption{\small{Convergence behavior of the achievable sum rate with Algorithm~\ref{alg:OP_Overall_Distr_Alg} for different transmit power levels $P^{\max}$, considering $M=64$ unit elements for each one of the deployed BDRISs. %Both cooperation and non-cooperation schemes ($\vect{\Pi}=\vect{0}$) were considered.
    }}
	\label{fig:Conv_DGC}
\end{figure}

In the performance results that follow, we have set the transmit power $P_b^{\max}=P^{\max}$ $\forall b\in\mathscr{B}$ and the noise variance $\sigma_{u,k}^2 = \sigma^2=-80$ dBm at all UEs. Also, the carrier frequency was set as $f_c=2.4$ GHz, the bandwidth as $\rm{BW}=300$ MHz, and $K=32$ SCs, unless otherwise specified, with the central frequency of each $k$-th SC defined as $f_k\triangleq f_c + (k - \frac{K+1}{2})\frac{\rm{BW}}{K}$ $\forall k=1,2,\dots,K$. The circuital elements for each BDRIS were set as follows: $L_1=2.5$ nH, $L_2=0.7$ nH, $R_0=1$ $\Omega$, $\psi_0=\frac{1}{50}$ S, $C_{\min}=0.2$ pF, and $C_{\max}=3$ pF. Finally, we have used the algorithmic parameters: $\tau=10^{-2}$ and $\epsilon = 10^{-3}$ (convergence threshold), while, for the step sizes $\rho^t$ and $\alpha^t$, we have used the updating rules $\alpha^t=\frac{1}{(t+2)^{0.61}}$ and $\rho^t = \frac{1}{(t+2)^{0.6}}$ that satisfy the requirements of Theorem~\ref{thm:conv_consensus}. All average performance metrics have been obtained using $100$ independent realizations of all involved wireless channels. %Two versions of the sum rate are included: \textit{i}) the ``Achievable Sum Rate,'' as defined in $\mathcal{OP}$'s objective function; and \textit{ii}) the ``Instantaneous Sum Rate'' \textcolor{red}{indicating the instantaneous sum-rate performance}. 
\begin{figure}[!t]
	\centering
	\includegraphics[trim=0cm 0cm 0cm 0.5cm,clip,width=3.75in]{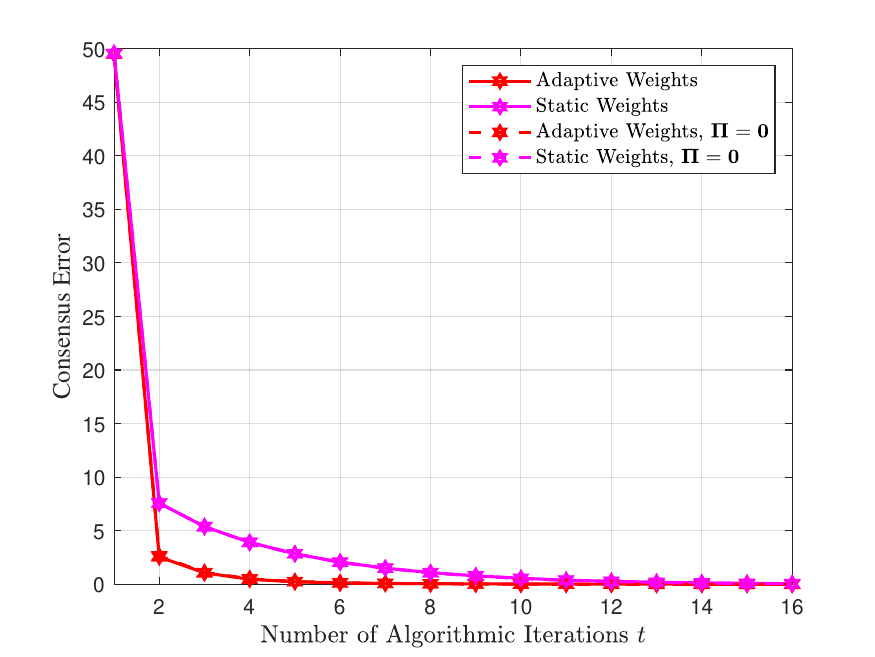}
	\caption{\small{Consensus error for the BDRIS parameters $\widetilde{\vect{C}}$ and $\widetilde{\vect{Q}}_p$ with respect to the number of iterations in Algorithm~\ref{alg:OP_Overall_Distr_Alg}, considering both static and adaptive weights for the network graph matrix $\vect{V}_{\rm net}$, $P^{\max} = 35$ dBm, and $M=64$ unit elements per metasurface.}}
	\label{fig:Cons_Error}
\end{figure}

\subsection{Behavior of Algorithm~\ref{alg:OP_Overall_Distr_Alg}} \label{sec:Performance}
We first examine the convergence of the proposed robust decentralized cooperative approach summarized in Algorithm~\ref{alg:OP_Overall_Distr_Alg} for different values of the maximum allowable transmit power $P^{\max}$ at each BS. Both cooperation and non-cooperation ($\vect{\Pi}=\vect{0}$; see \cite{JSTSP_distributed} for details) cases among the deployed BSs were considered. It can be observed in Fig.~\ref{fig:Conv_DGC} that, for cooperation scenarios, more iterations are needed to achieve convergence as the transmit power increases, leading to at most $T_{\max} \approx 25$ iterations when $P^{\max} = 35$ dBm, while convergence is achieved within $T_{\max} = 10$ iterations for $P^{\max} = 25$ dBm. However, all non-cooperative schemes converge within $T_{\max} = 15$ iterations for all transmit power levels, resulting in lower sum-rate performance as $P^{\max}$ increases. As a consequence, the proposed decentralized cooperative with minimal cooperation overhead is able to converge within a relatively small number of algorithmic iterations. 

Next, we investigate the performance of Algorithm~\ref{alg:OP_Overall_Distr_Alg} in terms of the consensus error measured, as presented in Theorem~\ref{thm:conv_consensus}, for two different cases related to the weights of the network graph $\vect{V}_{\rm net}$. In particular, in Fig.~\ref{fig:Cons_Error}, we depict the consensus error for the proposed design considering both static (i.e., $\vect{V}_{\rm net}^t = \vect{V}_{\rm net}$ $\forall t = 1,2,\dots,T_{\max}$) and adaptive weights, according to the proposed heuristic method detailed in Section~\ref{sec:V_net_Adaptive}. It is depicted that the consensus error is large enough in the first iterations since all BDRIS-involved design variables are initialized to be different among the BSs. Nevertheless, the consensus error disappears rapidly and tends to zero after $5$ iterations when considering the proposed approach with adaptive weights, while, when $\vect{V}_{\rm net}$ is left invariant, the consensus error decreases close to zero within $12$ iterations. This behavior remains unchanged for both cooperative and non-cooperative casess indicating that consensus is achieved rapidly among BSs, especially for the proposed approach regarding $\vect{V}_{\rm net}^t$'s evolution. More importantly, we have witnessed that the sum-rate performance does not change with the selection of the adaptive over static weights, since the achievable sum rate remains the same for both cases.
\begin{figure}[!t]
	\centering
	\includegraphics[trim=0cm 0cm 0cm 0.5cm,clip,width=3.75in]{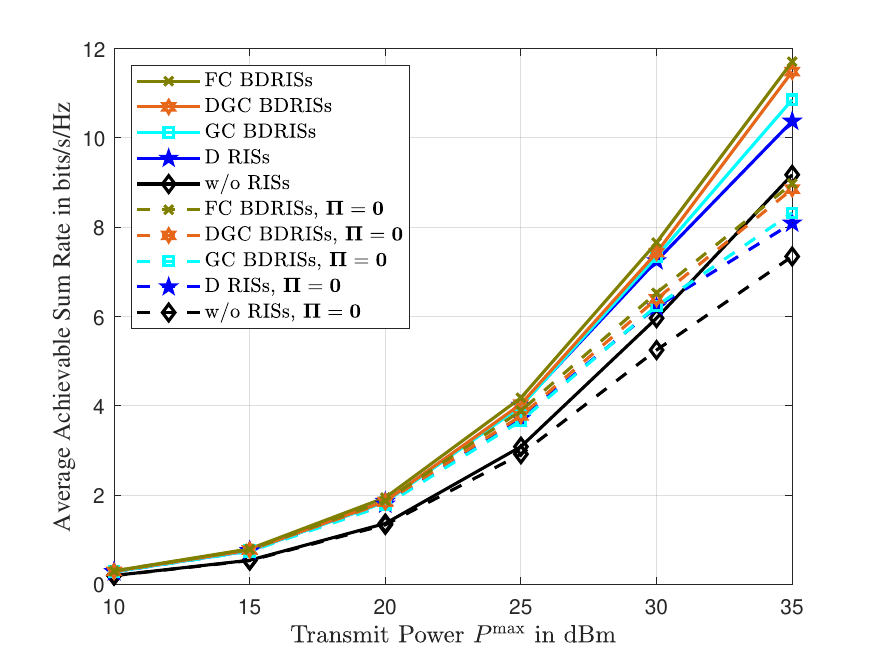}
	\caption{\small{Average achievable sum-rate performance versus the transmit power $P^{\max}$ of each BS, considering $R=2$ BDRISs of $M=64$ unit elements with different architectures.}}
	\label{fig:Rates_Power}
\end{figure}

\subsection{Sum-Rate Performance Evaluation} \label{sec:Benchmarks}
For comparison purposes, we have evaluated the average achievable sum rate with Algorithm~\ref{alg:OP_Overall_Distr_Alg} considering overall four BDRIS architectures, all resulting from the presented DGC in Section~\ref{sec:RIS_Response}. In particular, we have simulated: \textit{i}) the fully-connected BDRIS architecture with $G = 1$ and $\vect{Q}_p$ fixed as $\vect{Q}_p=\vect{I}_M$, which serves as an upper bound in terms of performance; \textit{ii}) the common Group-Connected (GC) architecture for $\vect{Q}_p=\vect{I}_M$; and \textit{iii}) the single-connected architecture (i.e., with the common diagonal response matrix) letting $G=M$ and $\vect{Q}_p=\vect{I}_M$, termed, henceforth, as ``FC BDRISs,'' ``GC BDRISs,'' and ``D RISs,'' respectively. The proposed design is termed as ``DGC BDRISs.'' For this comparison, we have set $\tau_{\rm FC}=10^{-2}$, $\tau_{\rm GC}=4\times10^{-2}$, and $\tau_{\rm D} = 5\times10^{-2}$.  In Fig.~\ref{fig:Rates_Power}, we depict the achievable sum-rate performance $\mathcal{R}$ as a function of the transmit power $P^{\max}$ of each BS considering both cooperation and non-cooperation cases. For all illustrated schemes, it is evident that the sum rates follow a non-decreasing trend as $P^{\max}$ increases. In addition, it is shown that the cooperation schemes outperform the corresponding non-cooperative ones, especially when $P^{\max}$ is larger than $25$ dBm, i.e., in the high Signal-to-Interference-and-Noise-Ratio (SINR) regime. Moreover, it can be observed that the ``FC BDRISs'' case outperforms all other architectures as anticipated, while the proposed performance with the ``DGC BDRISs'' architecture is very close to the former's one, achieving higher performance than the ``GC BDRISs'' and ``D RISs'' architectures when $P^{\max} \geq 30$ dBm. 
\begin{figure}[!t]
	\centering
	\includegraphics[trim=0cm 0cm 0cm 0.5cm,clip,width=3.75in]{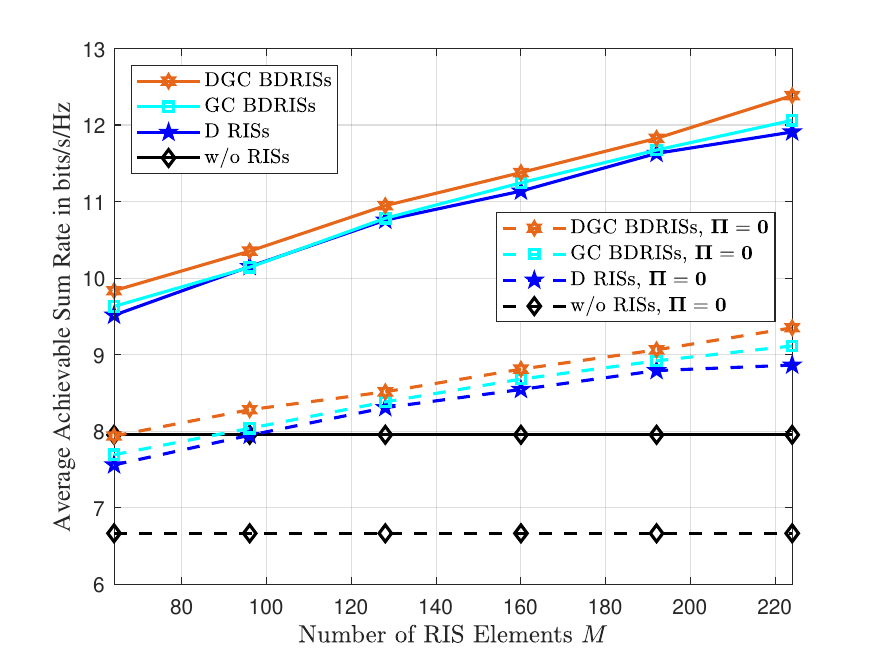}
	\caption{\small{Average achievable sum rate as a function of the common number $M$ of unit elements at each BDRIS, considering the transmit power $P^{\max}=30$ dBm and different metasurface architectures.}}
	\label{fig:Rates_RIS_Elem}
\end{figure}

The impact of the varying size of the BDRISs on the average achievable sum rate is illustrated in Fig.~\ref{fig:Rates_RIS_Elem} for different interconnection architectures among their unit elements, omitting though the ``FC RISs'' architecture due to its prohibitively large complexity when increasing $M$. For this figure, we have considered $G=8$ groups for the two GC architectures, $K=16$ SCs, and $P^{\max}=30$ dBm. As expected, increasing $M$ results in improved sum-rate performance for all investigated schemes, since all exhibit a non-decreasing trend. This trend is in fact more pronounced for cooperative scenarios (i.e., for $\vect{\Pi}\neq \vect{0}$) and all considered architectures since, as observed, the improvement slope is larger than that of the corresponding non-cooperative cases. In particular, the rate of improvement for the case of ``DGC BDRISs'' with cooperation among the BSs is estimated as $1.6\%$, whereas, for the corresponding non-cooperative scheme, it is quite lower and equal to $0.88\%$ across the entire range that $M$ spans. The same behavior is witnessed for the remaining two architectures (i.e., ``GC BDRISs'' and ``D RISs'') since, as depicted, their rates increase in a parallel way with respect to the proposed architecture ``DGC BDRISs.''

Finally, we compare our robust decentralized cooperative design approach with centralized schemes from the recent literature. At this point, we note that, to the best of the authors' knowledge, there is no recent work dealing with multiple BDRISs in a cell-free MIMO system setting. As a result, we compared the proposed decentralized approach with the recent study \cite{wang2025efficient}, which deals with a single-stream wideband transmission model and proposes a centralized design scheme for ``D RISs.'' For a fair comparison with this centralized algorithm, we also considered single-stream transmission which, in this case, reduces to the same objective function with the design metric in \cite{wang2025efficient}. Moreover, this study considered perfect CSI availability in the CPU for all communication links, and dealt with imperfect CSI knowledge only in the performance investigation, adopting the same model as in our case, i.e., $\hat{g} = g + e$, in order to testify the robustness of their proposed algorithm. 
\begin{figure}[!t]
	\centering
	\includegraphics[trim=0cm 0cm 0cm 0.5cm,clip,width=3.75in]{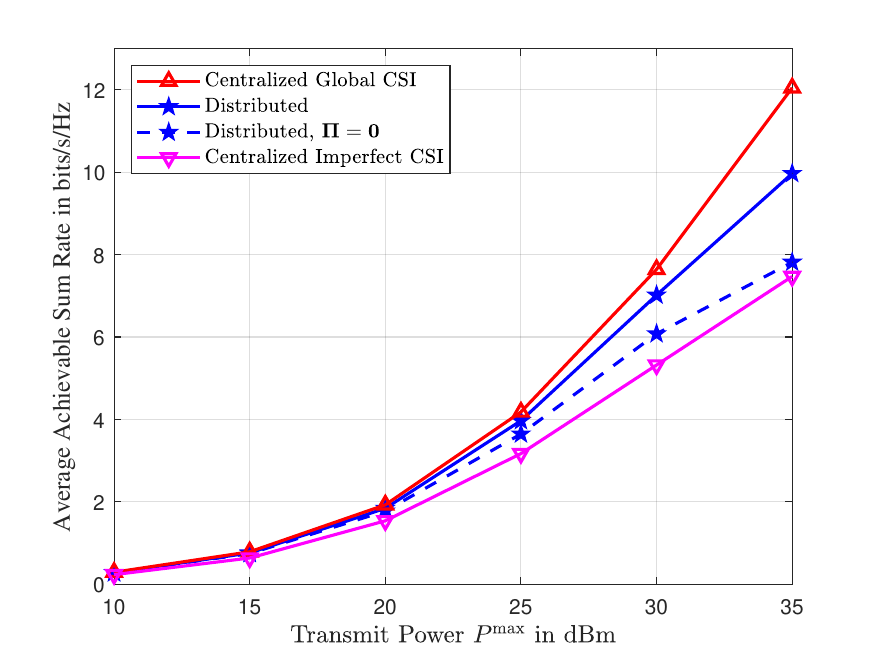}
	\caption{\small{Average achievable sum-rate performance with respect to the transmit power $P^{\max}$, considering BDRISs with $M=64$ unit elements as well as both global and imperfect CSI cases.}}
	\label{fig:Central_vs_Distrib}
\end{figure}
In Fig.~\ref{fig:Central_vs_Distrib}, we depict the average achievable sum rate for two centralized schemes: \textit{i}) the ``Centralized Global CSI,'' which clearly serves as an upper bound in terms of performance; and \textit{ii}) ``Centralized Imperfect CSI,'' which relies on $\hat{g}$ with fixed $\delta=0.2$, along with our ``D RISs'' schemes for both $\vect{\Pi}\neq \vect{0}$ and $\vect{\Pi} = \vect{0}$ cases, termed as ``Distributed'' and ``Distributed, $\vect{\Pi}=\vect{0}$,'' respectively. As illustrated, the proposed decentralized designs outperform the ``Centralized Imperfect CSI'' scheme for $P^{\max}\geq 20$ dBm, especially when cooperation among the BSs is enabled. Furthermore, the proposed decentralized cooperative scheme's performance is very close to that of the centralized upper bound up to $P^{\max} = 25$ dBm, while, in the high-SINR regime, i.e., for $P\geq 30$ dBm, there is a gap with maximum difference equal to $2$ bits/s/Hz between them. Evidently, the proposed algorithm is more robust than centralized designs that do not take into account imperfect CSI during the systems design process.

\section{Conclusions} \label{sec:Concl}
In this paper, we investigated a cell-free system comprising multiple multi-antenna BSs that perform OFDM transmissions towards multiple multi-antenna UEs under the assistance of multiple shared BDRISs, and presented a novel decentralized cooperative beamforming framework aiming to maximize the sum rate under imperfect CSI availability. We considered the case in which the metamaterial response is dictated by a frequency-dependent profile and presented a wideband DGC architecture for each deployed BDRIS. The joint design objective under consideration included the digital precoders at the multiple BSs as well as the tunable capacitances and permutation matrices at the multiple BDRISs. In addition, the proposed algorithmic approach accounted for the fact that the BDRISs are shared among the deployed BSs, incorporating inherent consensus-based updates along with minimal cooperation exchange messages during the overall system design process. Our extensive performance evaluation results demonstrated that the considered BDRIS architecture yields higher sum-rate performance than that of conventional diagonal and GC architectures, while its performance is close to that of fully-connected BDRISs. Moreover, it was showcased that the proposed decentralized consensus-based algorithm is capable of achieving superior sum-rate performance than benchmark centralized schemes with imperfect CSI, even in the case of non-cooperative beamforming, while it approaches that of perfect CSI designs when a CPU coordinates beamforming.

%----------------------------------------------------------%
%                      APPENDICES                          %
%----------------------------------------------------------%
\appendices
\section{Proof of Lemma~\ref{lem:surrogate_precoder}} \label{apx:surrogate_precoder_proof}
We start by noting that $\mathcal{R}_{u,k}$ can be equivalently rewritten, after invoking the matrix inversion lemma and using the definitions for matrices $\vect{T}_{u,k}$ and $\vect{L}_{u,k}$, as follows:
\begin{align}
    \mathcal{R}_{u,k}(\vect{S}_{u,u,k},\vect{T}_{u,k}) &= \!-\!\log_2\left| \vect{I}_{N_s}\!-\!\vect{S}_{u,u,k}^{\rm H}\vect{T}_{u,k}^{-1}\vect{S}_{u,u,k} \right|,
\end{align}
Then, according to \cite[Lemma 3]{wang2016design}, $\mathcal{R}_{u,k}$ is convex in $\{\vect{S}_{u,u,k},\vect{T}_{u,k}\}$ and can be lower bounded by its first-order Taylor expansion ($\breve{\mathcal{\mathcal{R}}}_{u,k}$) around matrices $\vect{S}_{u,u,k}$ and $\vect{T}_{u,k}$. Hence, to obtain the minorization function $\breve{\mathcal{R}}_{u,k}$, it suffices to compute the gradients $\nabla_{\vect{S}_{u,u,k}}\mathcal{R}_{u,k}$, $\nabla_{\vect{S}_{u,u,k}^*}$, and $\nabla_{\vect{T}_{u,k}}\mathcal{R}_{u,k}$. To this end, we use the differential rule ${\rm d}\mathcal{R}_{u,k}=-\frac{1}{\ln(2)}\trace(\vect{L}_{u,k}^{-1}{\rm d\vect{L}_{u,k}})$, whose application leads to the following gradient matrices:
\begin{align}
    \nabla_{\vect{S}_{u,u,k}}\mathcal{R}_{u,k} &= \frac{1}{\ln(2)}\vect{T}_{u,k}^{-\rm T}\vect{S}_{u,u,k}^*\vect{L}_{u,k}^{-\rm T},\\
    \nabla_{\vect{S}_{u,u,k}^*}\mathcal{R}_{u,k} &= \frac{1}{\ln(2)}\vect{T}_{u,k}^{-1}\vect{S}_{u,u,k}\vect{L}_{u,k}^{-1},\\
    \nabla_{\vect{T}_{u,k}}\mathcal{R}_{u,k} &= -\frac{1}{\ln(2)}\vect{T}_{u,k}^{-\rm T}\vect{S}_{u,u,k}^*\vect{L}_{u,k}^{-\rm T}\vect{S}_{u,u,k}^{\rm T}\vect{T}_{u,k}^{-\rm T}.
\end{align}
For the latter derivation the differential rule ${\rm d}\vect{Z}^{-1}= \vect{Z}^{-1}({\rm d}\vect{Z})\vect{Z}^{-1}$ was used. Next, it suffices to replace the derived gradient matrices in the Taylor formula properly in order for $\vect{W}_{b,u,k}$ to appear, which after som straightforward algebraic manipulations and usage of trace's cyclic property, as well as the identity $\Re\{\vect{X}\} = \frac{1}{2}(\vect{X} + \vect{X}^*)$ for an arbitrary complex matrix $\vect{X}$, leads to the desired expression for the surrogate $\breve{\mathcal{R}}_{u,k}$. The latter concludes the proof.

\section{Proof of Theorem~\ref{thm:gradient_Capacitors}} \label{apx:gradient_Capacitors_proof}
To prove this theorem, it suffices to derive the gradient of $\mathcal{R}_{u,k}$ with respect to the $g$-th group of the capacitor matrix $\vect{C}_{b,r}$, whose expression ($\nabla_{\vect{C}_{b,r,g}}\mathcal{R}_{u,k}$) in \eqref{eqn:Cap_gradient} can be acquired by first noticing that the channel matrix $\widetilde{\vect{H}}_{b,u,k}$ can be equivalently expressed as follows:
\begin{align}
    &\widetilde{\vect{H}}_{b,u,k} = \vect{H}_{b,u,k} + \sum_{r=1}^R\vect{G}_{r,u,k}\vect{Q}_{p,b,r}\vect{\Phi}_{r,k}\vect{Q}_{p,b,r}^{\rm T}\vect{F}_{b,r,k} \label{eqn:channel_unfolded}\\
    &= \vect{H}_{b,u,k} + \sum_{r=1}^R\sum_{g=1}^G\vect{G}_{r,u,k,g}\vect{Q}_{p,b,r,g}\vect{\Phi}_{r,k,g}\vect{Q}_{p,b,r,g}^{\rm T}\vect{F}_{b,r,k,g}, \nonumber
\end{align}
where the definitions in \eqref{eqn:submatrix_1}--\eqref{eqn:submatrix_3} have been used. Then, we also note that $\vect{C}_{b,r,g}$ is encapsulated in $\vect{\Phi}_{r,k,g}$ through the matrix functions expressed in \eqref{eqn:RIS_Scattering} and \eqref{eqn:impedances}, which implies that it suffices to properly apply the chain rule for the function $\mathcal{R}_{u,k}$ to proceed. However, in the rate expression $\mathcal{R}_{u,k}$ (see, e.g., the $k$-th summand of \eqref{eqn:sum_rate_per_user}), $\widetilde{\vect{H}}_{b,u,k}$ appears in both the matrices $\vect{S}_{u,u,k}$ and $\vect{P}_{u,k}^{-1}$, as well as in $\vect{S}_{u,u,k}^{\rm H}$. For brevity and to avoid redundancy, we focus on the derivation related to $\vect{S}_{u,u,k}$ and partially to $\vect{P}_{u,k}^{-1}$. We also adopt the notation $\mathcal{D}_{\vect{Z}}\vect{F}$ introduced in \cite[Definition 3.1]{hjorungnes2011complex}, to represent the Jacobian matrix of the matrix function $\vect{F}(\vect{Z},\vect{Z}^*):\mathbb{C}^{I\times J}\times\mathbb{C}^{I\times J}\rightarrow\mathbb{C}^{Q\times P}$ with respect to the matrix variable $\vect{Z}\in\mathbb{C}^{I\times J}$. Then, it is deduced that $\nabla_{\vect{Z}}\vect{F} = (\mathcal{D}_{\vect{Z}^*}\vect{F})^{\rm T} \in\mathbb{C}^{IJ\times QP}$. 

Using the differential rule ${\rm d}\ln(|\vect{Z}|) = \trace(\vect{Z}^{-1}{\rm d\vect{Z}})$ along with the identity $\trace(\vect{A}^{\rm T}\vect{B}) = \operatorname{vec}^{\rm T}(\vect{A})\operatorname{vec}(\vect{B})$, the following Jacobian is obtained: 
\begin{equation} \label{eqn:grad_cap_part1}
    \mathcal{D}_{\vect{S}_{u,u,k}}\mathcal{R}_{u,k} = \frac{1}{\ln(2)}\operatorname{vec}^{\rm T}\left(\vect{P}_{u,k}^{-\rm T}\vect{S}_{u,u,k}^*\vect{K}_{u,k}^{-\rm T}\right).
\end{equation}
Next, we compute $\mathcal{D}_{\vect{\Phi}_{r,k,g}}\vect{S}_{u,u,k}$ noting that $\vect{S}_{u,u,k} = \widetilde{\vect{H}}_{b,u,k}\vect{W}_{b,u,k} + \vect{R}_{u,k}$, where the latter matrix is independent of $\vect{C}_{b,r,g}$ and is defined in the Lemma~\ref{lem:surrogate_precoder}. It can be shown that:
\begin{equation} \label{eqn:grad_cap_part2}
    \mathcal{D}_{\vect{\Phi}_{r,k,g}}\vect{S}_{u,u,k} = \vect{\Delta}_{b,r,u,k,g}^*\otimes\vect{\Xi}_{b,r,u,k,g}^*,
\end{equation}
which can be derived by unfolding the matrix $\widetilde{\vect{H}}_{b,u,k}$, as explained above, and using the property $\operatorname{vec}(\vect{A}\vect{B}\vect{C}) = (\vect{C}^{\rm T}\otimes\vect{A})\operatorname{vec}(\vect{B})$ together with the matrix definitions \eqref{eqn:Delta_matrix} and \eqref{eqn:Xi_matrix}. 

To proceed with the derivative for $\mathcal{D}_{\vect{A}_{r,k,g}}\vect{\Phi}_{r,k,g}$, $\vect{\Phi}_{r,k,g}=\vect{X}_{r,k,g}^{-1}(\vect{X}_{r,k,g}-2\vect{A}_{r,k,g})$ holds due to \eqref{eqn:RIS_Scattering}, where $\vect{X}_{r,k,g}$ is defined in \eqref{eqn:X_matrix}. Then, the following is deduced:
\begin{align*} \label{eqn:differntial_X}
    {\rm d}\vect{\Phi}_{r,k,g} &= ({\rm d}\vect{X}_{r,k,g}^{-1})(\vect{X}_{r,k,g}-2\vect{A}_{r,k,g}) - \vect{X}_{r,k,g}^{-1}({\rm d}\vect{A}_{r,k,g}) \\
    &\begin{aligned}
        &\stackrel{(a)}{=}-\vect{X}_{r,k,g}^{-1}({\rm d}\vect{A}_{r,k,g})\vect{X}_{r,k,g}^{-1}(\vect{X}_{r,k,g}-2\vect{A}_{r,k,g})\\ 
        &\quad-\vect{X}_{r,k,g}^{-1}({\rm d}\vect{A}_{r,k,g}),
    \end{aligned}    
\end{align*}
where $(a)$ follows from \cite[Proposition 3.8]{hjorungnes2011complex} and due to the fact that ${\rm d}\vect{X}_{r,k,g} = {\rm d}\vect{A}_{r,k,g}$. The differential above can be also expressed in vectorized form as follows:
\begin{equation}
    {\rm d}\operatorname{vec}(\vect{\Phi}_{r,k,g}) = -2\psi_0\left( \vect{X}_{r,k,g}^{-1}\otimes\vect{X}_{r,k,g}^{-1} \right){\rm d}\operatorname{vec}(\vect{A}_{r,k,g}),
\end{equation}
where we have again used the property for $\operatorname{vec}(\vect{A}\vect{B}\vect{C})$ and the fact that $\vect{X}_{r,k,g}=\vect{X}_{r,k,g}^{\rm T}$ which is valid for feasibility reasons of the admittance matrix $\vect{A}_{r,k,g}$. Then, according to \cite[Table 3.2]{hjorungnes2011complex}, it can be concluded that:
\begin{equation} \label{eqn:grad_cap_part3}
    \mathcal{D}_{\vect{A}_{r,k,g}}\vect{\Phi}_{r,k,g} = -2\psi_0\left(\vect{X}_{r,k,g}^{-1}\otimes\vect{X}_{r,k,g}^{-1}\right).
\end{equation}

To compute $\mathcal{D}_{\vect{C}_{b,r,g}}\vect{A}_{r,k,g}$, it is noted that, due to~\eqref{eqn:impedances}, each diagonal entry in the matrix $\vect{A}_{r,k,g}$ depends on at least one entry of $\vect{C}_{b,r,g}$, while the non-diagonal entries are affected only by a single entry of the corresponding matrix. Therefore, it suffices to compute the partial derivatives $\frac{\partial\operatorname{vec}(\vect{A}_{r,k,g})}{\partial\operatorname{vec}(\vect{C}_{b,r,g})^{\rm T}}$, entry by entry, resulting in the expression~\eqref{eqn:Cap_gradient_part2}. Consequently, the chain rule yields that $\mathcal{D}_{\vect{C}_{b,r,g}}\mathcal{R}_{u,k}$ is equal to the product of the quantities in~\eqref{eqn:grad_cap_part1}, \eqref{eqn:grad_cap_part2}, and \eqref{eqn:grad_cap_part3} as well as $\mathcal{D}_{\vect{C}_{b,r,g}}\vect{A}_{r,k,g}$. Similarly, the derivative $\mathcal{D}_{\vect{C}_{b,r,g}}\mathcal{R}_{u,k}$ through $\vect{S}_{u,u,k}^{\rm H}$ can be readily obtained, and it can also be observed that there is symmetry between the latter two terms, in the sense that the first is the conjugate of the other. This holds because $\mathcal{R}_{u,k}$ and $\vect{C}_{b,r,g}$ are real entities. Moreover, using the definition of $\vect{P}_{u,k} = \sum_{q=1,q\neq u}^U \vect{S}_{u,q,k} \vect{S}_{u,q,k}^{\rm H} + \sigma_{u,k}^2\vect{I}_{N_r}$, the dependence on $\vect{C}_{b,r,g}$ is expressed through the matrices $\vect{S}_{u,q,k}$ and $\vect{S}_{u,q,k}^{\rm H}$. Then, the following derivation is deduced:
\begin{equation} \label{eqn:differential_rate_P}
    {\rm d}\mathcal{R}_{u,k} = \frac{-1}{\ln(2)}\trace\left( \vect{K}_{u,k}^{-1}\vect{S}_{u,u,k}^{\rm H}\vect{P}_{u,k}^{-1}({\rm d}\vect{P}_{u,k})\vect{P}_{u,k}^{-1}\vect{S}_{u,u,k} \right),
\end{equation}
where, invoking the linearity and multiplication properties for the complex matrix differential, yields:
\begin{equation} \label{eqn:differential_cap_P}
    {\rm d}\vect{P}_{u,k} = \sum_{\substack{q=1,\\q\neq u}}^U \left( ({\rm d}\vect{S}_{u,q,k})\vect{S}_{u,q,k}^{\rm H} + \vect{S}_{u,q,k}({\rm d}\vect{S}_{u,q,k}^{\rm H}) \right).
\end{equation}
Hence, it suffices to substitute \eqref{eqn:differential_cap_P} into \eqref{eqn:differential_rate_P} and repeat the previous derivations for both $\vect{S}_{u,q,k}$ and $\vect{S}_{u,q,k}^{\rm H}$ to obtain the remaining derivatives. Finally, combining all above and after some  straightforward algebraic manipulations (omitted due to space limitations), along with the usage of the mixed product property $(\vect{A}\otimes\vect{C})(\vect{B}\otimes\vect{D})=(\vect{A}\vect{B})\otimes(\vect{C}\vect{D})$, the matrix $\vect{\Theta}_{b,r,u,k,g}$ defined in \eqref{eqn:Cap_gradient_part1} appears; this completes the proof.

\bibliographystyle{IEEEtran}
\bibliography{References_Journal}

@inproceedings{Katsanos_Cell_Free_conf,
    author = {K. D. Katsanos and G. C. Alexandropoulos},
    title = {Robust consensus–based distributed beamforming for wideband cell-free multi-{RIS} {MISO} systems},
    booktitle = {Proc. IEEE Asilomar Signals, Sys., Comp. Conf.},
    address = {Pacific Grove, USA},
    year = {2025}
}

@article{JSTSP_distributed_all,
  author={Katsanos, Konstantinos D. and Lorenzo, Paolo Di and Alexandropoulos, George C.},
  journal={IEEE J. Sel. Topics Signal Process.}, 
  title={Multi-{RIS}-Empowered Multiple Access: A Distributed Sum-Rate Maximization Approach},
  year={2024},
  volume={18},
  number={7},
  pages={1324-1338}
}

@article{JSTSP_distributed,
  author={Katsanos, Konstantinos D. and others},
  journal={IEEE J. Sel. Topics Signal Process.}, 
  title={Multi-{RIS}-Empowered Multiple Access: A Distributed Sum-Rate Maximization Approach},
  year={2024},
  volume={18},
  number={7},
  pages={1324-1338}
}

@inproceedings{IBC_distributed2024,
  author={Katsanos, Konstantinos D. and others},
  booktitle={Proc. IEEE SPAWC},  
  title={The Interference Broadcast Channel with Reconfigurable Intelligent Surfaces: A Cooperative Sum-Rate Maximization Approach}, 
  address={Lucca, Italy},
  year={2024}
}

@article{ngo2017cell,
  title={Cell-free massive {MIMO} versus small cells},
  author={Ngo, Hien Quoc and others},
  journal={IEEE Trans. Wireless Commun.},
  volume={16},
  number={3},
  pages={1834--1850},
  year={2017},
  publisher={IEEE}
}

@article{ma2022cooperative,
  title={Cooperative beamforming design for multiple {RIS}-assisted communication systems},
  author={Ma, Xiaoyan and others},
  journal={IEEE Trans. Wireless Commun.},
  volume={21},
  number={12},
  pages={10949--10963},
  year={2022},
  publisher={IEEE}
}

@article{al2024performance,
  title={Performance of multi-{RIS}-aided cell-free massive {MIMO}: Do more {RIS}s always help?},
  author={Al-Nahhas, Bayan and others},
  journal={IEEE Trans. Commun.},
  volume={72},
  number={7},
  pages={4319--4332},
  year={2024},
  publisher={IEEE}
}

@article{huang2020decentralized,
  title={Decentralized beamforming design for intelligent reflecting surface-enhanced cell-free networks},
  author={Huang, Shaocheng and others},
  journal={IEEE Wireless Commun. Lett.},
  volume={10},
  number={3},
  pages={673--677},
  year={2020},
  publisher={IEEE}
}

@article{xu2023algorithm,
  title={Algorithm-unrolling-based distributed optimization for {RIS}-assisted cell-free networks},
  author={Xu, Wangyang and others},
  journal={IEEE Internet Things J.},
  volume={11},
  number={1},
  pages={944--957},
  year={2023},
  publisher={IEEE}
}

@article{ni2022partially,
  title={Partially distributed beamforming design for {RIS}-aided cell-free networks},
  author={Ni, Pengfei and others},
  journal={IEEE Trans. Veh. Technol.},
  volume={71},
  number={12},
  pages={13377--13381},
  year={2022},
  publisher={IEEE}
}

@article{yang2016parallel,
  title={A parallel decomposition method for nonconvex stochastic multi-agent optimization problems},
  author={Yang, Yang and others},
  journal={IEEE Trans. Signal Process.},
  volume={64},
  number={11},
  pages={2949--2964},
  year={2016},
  publisher={IEEE}
}

@article{Alexandropoulos2023_RISDeployment,
  author  = {Alexandropoulos, George~C. and others},
  title   = {{RIS}-Enabled Smart Wireless Environments: Deployment Scenarios, Network Architecture, Bandwidth, and Area of Influence},
  journal = {EURASIP J. Wireless Commun. Netw.},
  year    = {2023},
  volume  = {2023},
  number  = {103}
}

@article{xin2020general,
  title={A general framework for decentralized optimization with first-order methods},
  author={Xin, Ran and others},
  journal={Proc. IEEE},
  volume={108},
  number={11},
  pages={1869--1889},
  year={2020},
  publisher={IEEE}
}

@article{lorenzo2016next,
  title={{NEXT}: In-network nonconvex optimization},
  author={Di Lorenzo, Paolo and Scutari, Gesualdo},
  journal={IEEE Trans. Signal Inf. Process. Netw.},
  volume={2},
  number={2},
  pages={120--136},
  year={2016},
  publisher={IEEE}
}

@article{zhu2010discrete,
  title={Discrete-time dynamic average consensus},
  author={Zhu, Minghui and Mart{\'\i}nez, Sonia},
  journal={Automatica},
  volume={46},
  number={2},
  pages={322--329},
  year={2010},
  publisher={Elsevier}
}

@article{zhang2021joint,
  title={A joint precoding framework for wideband reconfigurable intelligent surface-aided cell-free network},
  author={Zhang, Zijian and Dai, Linglong},
  journal={IEEE Trans. Signal Process.},
  volume={69},
  pages={4085--4101},
  year={2021},
  publisher={IEEE}
}

@article{wang2025efficient,
  title={Efficient Joint Precoding Design for Wideband Intelligent Reflecting Surface-Assisted Cell-Free Network},
  author={Wang, Yajun and others},
  journal={IEEE Trans. Commun.},
  year={2025},
  note={early access},
  publisher={IEEE}
}

@article{nerini2024static,
  title={Static grouping strategy design for beyond diagonal reconfigurable intelligent surfaces},
  author={Nerini, Matteo and others},
  journal={IEEE Commun. Lett.},
  volume={28},
  number={7},
  pages={1708--1712},
  year={2024},
  publisher={IEEE}
}

@book{bauschke2011convex,
  title={Convex analysis and monotone operator theory in {Hilbert} spaces},
  author={Bauschke, HH},
  year={2011},
  publisher={Springer-Verlag}
}

@book{dattorro2010convex,
  title={Convex optimization \& {Euclidean} distance geometry},
  author={Dattorro, Jon},
  year={2010},
  publisher={Meboo Publishing}
}

@book{Boyd_2004,
	author={Boyd, S.P. and Vandenberghe, L.},
    title={Convex optimization},  	
    isbn={9780521833783},
    lccn={03063284},
    year={2004},
    publisher={Cambridge University Press}
}

@book{hjorungnes2011complex,
  title={Complex-valued matrix derivatives: with applications in signal processing and communications},
  author={Hj{\o}rungnes, Are},
  year={2011},
  publisher={Cambridge University Press}
}

@article{duff2001algorithms,
  title={On algorithms for permuting large entries to the diagonal of a sparse matrix},
  author={Duff, Iain S and Koster, Jacko},
  journal={SIAM J. Matrix Analysis Appl.},
  volume={22},
  number={4},
  pages={973--996},
  year={2001},
  publisher={SIAM}
}

@article{wang2016design,
  title={Design of {PAR}-constrained sequences for {MIMO} channel estimation via majorization--minimization},
  author={Wang, Zhongju and others},
  journal={IEEE Trans. Signal Process.},
  volume={64},
  number={23},
  pages={6132--6144},
  year={2016},
  publisher={IEEE}
}

@article{li2023dynamic,
  title={A dynamic grouping strategy for beyond diagonal reconfigurable intelligent surfaces with hybrid transmitting and reflecting mode},
  author={Li, Hongyu and others},
  journal={IEEE Trans. Veh. Technol.},
  volume={72},
  number={12},
  pages={16748--16753},
  year={2023},
  publisher={IEEE}
}

@article{nerini2024universal,
  title={A universal framework for multiport network analysis of reconfigurable intelligent surfaces},
  author={Nerini, Matteo and others},
  journal={IEEE Trans. Wireless Commun.},
  volume={23},
  number={10},
  pages={14575--14590},
  year={2024},
  publisher={IEEE}
}

@book{Zhang_2017,
	author={Zhang, Xian-Da},
	title={Matrix Analysis and Applications},
	year={2017},
	publisher={Cambridge University Press}
}

@article{li2025tutorial,
  title={A Tutorial on Beyond-Diagonal Reconfigurable Intelligent Surfaces: Modeling, Architectures, System Design and Optimization, and Applications},
  author={Li, Hongyu and others},
  journal={arXiv preprint:2505.16504},
  year={2025}
}

@article{maraqa2025beyond,
  title={Beyond diagonal {RIS}-aided wireless communications systems: State-of-the-art and future research directions},
  author={Maraqa, Omar and others},
  journal={arXiv preprint:2503.08826},
  year={2025}
}

@article{shen2021modeling,
  title={Modeling and architecture design of reconfigurable intelligent surfaces using scattering parameter network analysis},
  author={Shen, Shanpu and others},
  journal={IEEE Trans. Wireless Commun.},
  volume={21},
  number={2},
  pages={1229--1243},
  year={2021},
  publisher={IEEE}
}

@article{LSN+23,
  title={Reconfigurable intelligent surfaces 2.0: Beyond diagonal phase shift matrices},
  author={Li, Hongyu and others},
  journal={IEEE Commun. Mag.},
  volume = {62}, 
  number = {3},
  pages={102--108},
  year={2023},
  publisher={IEEE}
}

@article{li2025beamforming,
  title={Beamforming design for beyond diagonal {RIS}-aided cell-free massive {MIMO} systems},
  author={Li, Yizhuo and others},
  journal={arXiv preprint:2503.07189},
  year={2025}
}

@article{hua2025cell,
  title={Cell-free massive {MIMO} {SWIPT} with beyond diagonal reconfigurable intelligent surfaces},
  author={Hua, Thien Duc and others},
  journal={IEEE Trans. Commun.},
  year={early access, 2025},
  publisher={IEEE}
}

@inproceedings{santamaria2024mimo,
  title={{MIMO} capacity maximization with beyond-diagonal {RIS}},
  author={Santamaria, Ignacio and others},
  booktitle={Proc. IEEE SPAWC},
  address={Lucca, Italy},
  pages={936--940},
  year={2024}
}

@article{zhou2023optimizing,
  title={Optimizing power consumption, energy efficiency, and sum-rate using beyond diagonal {RIS}—A unified approach},
  author={Zhou, Yuyan and others},
  journal={IEEE Trans. Wireless Commun.},
  volume={23},
  number={7},
  pages={7423--7438},
  year={2023},
  publisher={IEEE}
}

@article{gao2025multi,
  title={Multi-{RIS} Aided Multi-Cell Wireless Networks: Joint Beamforming Design Combined with Selection Strategy},
  author={Gao, Zilu and others},
  journal={IEEE Internet Things J.},
  volume={12},
  number={14},
  pages={27943-27956},
  year={2025},
  publisher={IEEE}
}

@article{lim2025distributed,
  title={Distributed Graph-Based Learning for User Association and Beamforming Design in Multi-{RIS} Multi-Cell Networks},
  author={Lim, Byungju and Vu, Mai},
  journal={IEEE Trans. Wireless Commun.},
  volume={24},
  number={7},
  pages={6118-6134},
  year={2025},
  publisher={IEEE}
}

@article{Basar2024_RIS_6G,
  author  = {E. Basar and others},
  title   = {Reconfigurable Intelligent Surfaces for 6{G}: {E}merging Hardware Architectures, Applications, and Open Challenges},
  journal = {IEEE Veh. Technol. Mag.},
  year    = {2024},
  volume  = {19},
  number  = {3},
  pages   = {27--47}
}

@article{Jian2022_RIS_Hardware,
  author  = {M.~Jian and others},
  title   = {Reconfigurable Intelligent Surfaces for Wireless Communications: {O}verview of Hardware Designs, Channel Models, and Estimation Techniques},
  journal = {Intell. Conv. Netw.},
  year    = {2022},
  volume  = {3},
  number  = {1},
  pages   = {1--32}
}

@article{HRIS_CE,
 author={Haiyang Zhang and others}, 
 title = {Channel estimation with hybrid reconfigurable intelligent metasurfaces},
 journal={IEEE Trans. Commun.},
 volume={71},
 number={4},
 pages={2441--2456},
 year = {2023}
}

@article{Swindlehurst_CE,
  author={Swindlehurst, A. Lee and others},
  title={Channel Estimation With Reconfigurable Intelligent Surfaces-- {A} General Framework},
  journal={Proc. IEEE},
  pages={1--27},
  year={2022}
}

@article{elhoushy2021cell,
  title={Cell-free massive {MIMO}: A survey},
  author={Elhoushy, Salah and others},
  journal={IEEE Commun. Surveys Tuts.},
  volume={24},
  number={1},
  pages={492--523},
  year={2021},
  publisher={IEEE}
}

\end{document}